\documentclass[twocolumn,showpacs,showkeys,preprintnumbers,pre,amssymb,aps]{revtex4} 

\usepackage{bm}
\usepackage{dcolumn}
\usepackage{graphicx}
\usepackage{psfrag}
\usepackage{subfigure}

\begin{document}

\title{
  Advancing density waves and phase transitions in a\\
  velocity dependent randomization traffic cellular automaton
}

\author{Sven Maerivoet}
  \email{sven.maerivoet@esat.kuleuven.ac.be}
\author{Bart De Moor}
  \email{bart.demoor@esat.kuleuven.ac.be}
\affiliation{
  Department of Electrical Engineering ESAT-SCD (SISTA)\footnote{
    Phone: +32 (0)16 32 17 09 Fax: +32 (0)16 32 19 70\\
    URL: \texttt{http://www.esat.kuleuven.ac.be/scd}
  }, Katholieke Universiteit Leuven\\
  Kasteelpark Arenberg 10, 3001 Leuven, Belgium
}

\date{\today}

\begin{abstract}
  Within the class of stochastic cellular automata models of traffic flows, we 
  look at the velocity dependent randomization variant (VDR-TCA) whose 
  parameters take on a specific set of extreme values. These initial conditions 
  lead us to the discovery of the emergence of four distinct phases. Studying 
  the transitions between these phases, allows us to establish a rigorous 
  classification based on their tempo-spatial behavioral characteristics. As a 
  result from the system's complex dynamics, its flow-density relation exhibits 
  a non-concave region in which forward propagating density waves are 
  encountered. All four phases furthermore share the common property that moving 
  vehicles can never increase their speed once the system has settled into an 
  equilibrium.
\end{abstract}



\preprint{SISTA Internal Report 03-111}

\maketitle


\section{Introduction}

In the field of traffic flow modeling, microscopic traffic simulation has always 
been regarded as a time consuming, complex process involving detailed models 
that describe the behavior of individual vehicles. Approximately a decade ago, 
however, new microscopic models were being developed, based on the cellular 
automata programming paradigm from statistical physics. The main advantage was 
an efficient and fast performance when used in computer simulations, due to 
their rather low accuracy on a microscopic scale. These so-called \emph{traffic 
cellular automata} (TCA) are discrete in nature, in the sense that time advances 
with discrete steps and space is coarse-grained (allowing for high-speed 
simulations, especially when they are performed on a platform for parallel 
computation) \cite{MAERIVOET:03d}.

The seminal work done by Nagel and Schreckenberg in the construction of their 
stochastic traffic cellular automaton (i.e., the STCA) \cite{NAGEL:92}, led to a 
global adoptation of the TCA modeling scheme. In order to generate traffic jams, 
their model needs some kind of fluctuations, which are introduced by means of a 
noise parameter. But one of the artefacts associated with this STCA model, is 
that it gives rise to (many) unstable traffic jams \cite{NAGEL:94b}. A possible 
approach to achieve stable traffic jams, is to reduce the outflow from such 
jams, which can be accomplished by implementing so-called \emph{slow-to-start 
behavior}. The name is derived from the fact that vehicles exiting jam fronts 
are obliged to wait a small amount of time. One of these implementations, is the 
approach initially followed by Takayasu and Takayasu \cite{TAKAYASU:93}. They 
made the stochastic noise dependent on the distance between two consecutive 
vehicles. Another implementation is based on making this noise parameter 
dependent on the velocity of a vehicle, leading to the development of the 
\emph{velocity dependent randomization} (VDR) TCA. This TCA model moreover 
exhibits metastability and hysteresis phenomena \cite{BARLOVIC:98}.

In this context, our paper addresses the VDR-TCA slow-to-start model whose 
parameters take on a specific set of extreme values. Even though this bears 
little direct relevance for the understanding of traffic flows, it will lead to 
an induced `anomalous' behavior and complex system dynamics, resulting in four 
distinct emergent phases. We study these on the basis of the system's relation 
between density and flow (i.e., the fundamental diagram) which exhibits a 
non-concave region where, in contrast to the properties of congested traffic 
flow, density waves are propagated \emph{forwards}. We furthermore investigate 
the system's tempo-spatial evolution in each of these four phases, and consider 
an order parameter that allows us to track the transitions between them. 

Although some related studies exist (e.g., 
\cite{AWAZU:99,GRABOLUS:01,GRAY:01,LEVEQUE:01,NISHINARI:03,ZHANG:03}), the 
special behavior dis\-cus\-sed in this paper has not been reported as such, as 
most previously done research is largely devoted to empirical and analytical 
discussions about these TCA models, \emph{operating under normal conditions}. To 
streng\-then our claims, we compare our results with the existing literature at 
the end of this paper.

In section \ref{section:ExperimentalSetup}, we briefly describe the concept of a 
traffic cellular automaton and the experimental setup used when performing the 
simulations. The selected VDR-TCA model is presented in section \ref{sec:VDR}, 
as well as an account of some behavioral characteristics under normal operating 
conditions. The results of our various experiments with more complex system 
dynamics are subsequently presented and extensively discussed in section 
\ref{section:Results}, after which the paper concludes with a comparison with 
existing literature in section \ref{section:LiteratureStudy} and a summary in 
section \ref{section:Summary}.

\section{Experimental setup}
\label{section:ExperimentalSetup}

In this section, we introduce the operational characteristics of a standard 
single-lane traffic cellular automaton model. To avoid confusion with some of 
the notations in existing literature, we explicitly state our definitions.

  \subsection{Geometrical description}

\begin{figure}[!ht]
  \centering
  \psfrag{t}[][]{\footnotesize{$t$}}
  \psfrag{t+1}[][]{\footnotesize{$t + 1$}}
  \psfrag{i}[][]{\footnotesize{$i$}}
  \psfrag{j}[][]{\footnotesize{$j$}}
  \psfrag{deltat}[][]{\footnotesize{$\Delta T$}}
  \psfrag{deltax}[][]{\footnotesize{$\Delta X$}}
  \psfrag{gsi}[][]{\footnotesize{$g_{s_{i}}$}}
  \includegraphics[width=8.2cm]{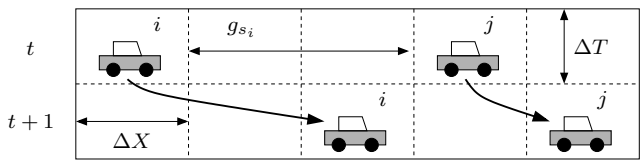}
  \caption{
    Schematic diagram of the operation of a single-lane traffic cellular 
    automaton (TCA); here, the time axis is oriented downwards, the space axis 
    extends to the right. The TCA's configuration is shown for two consecutive 
    time steps $t$ and $t + 1$, during which two vehicles $i$ and $j$ propagate 
    through the lattice. Without loss of generality, we denote the number of 
    empty cells in front of vehicle $i$ as its space gap $g_{s_{i}}$.
  }
  \label{fig:TCABasics}
\end{figure}

Let us describe the operation of a single-lane traffic cellular automaton as 
depicted in Fig.~\ref{fig:TCABasics}. We assume $N$ vehicles are driving on a 
circular lattice containing $K$ cells, i.e., periodic boundary conditions (each 
cell can be occupied by at most one vehicle at a time). Time and space are 
discretized, with $\Delta T = 1$~s and $\Delta X = 7.5$~m, leading to a velocity 
discretization of $\Delta V = 27$~km/h. Furthermore, the velocity $v_{i}$ of a 
vehicle $i$ is constrained to an integer in the range $\lbrace 0, \ldots, 
v_{\mbox{max}} \rbrace$, with $v_{\mbox{max}}$ typically 5~cells/s 
(corresponding to 135~km/h).

Each vehicle $i$ has a space headway $h_{s_{i}}$ and a time headway $h_{t_{i}}$, 
defined as follows:

\begin{eqnarray}
  h_{s_{i}} & = & L_{i} + g_{s_{i}}, \label{eq:SpaceHeadway}\\
  h_{t_{i}} & = & \rho_{i} + g_{t_{i}} \label{eq:TimeHeadway}.
\end{eqnarray}

In these definitions, $g_{s_{i}}$ and $g_{t_{i}}$ denote the space and time gaps 
respectively; $L_{i}$ is the length of a vehicle and $\rho_{i}$ is the occupancy 
time of the vehicle (i.e., the time it `spends' in one cell). Note that in a 
traffic cellular automaton the space headway of a vehicle is always an integer 
number, representing a multiple of the spatial discretisation $\Delta X$ in real 
world measurement units. So in a jam, it is taken to be equal to the space the 
vehicle occupies, i.e., $h_{s_{i}} =$~1~cell.

Local interactions between individual vehicles in a traffic stream are modeled 
by means of a rule set that reflects the car-following behavior of a cellular 
automaton evolving in time and space. In this paper, we assume that all vehicles 
have the same physical characteristics (i.e., homogeneous traffic). The system's 
state is changed through synchronous position updates of all the vehicles, based 
on a rule set that reflects the car-following behavior.\\

Most rule sets of TCA models do not use the space headway $h_{s_{i}}$ or the 
space gap $g_{s_{i}}$, but are instead based on the number of empty cells 
$d_{i}$ in front of a vehicle $i$. Keeping equation (\ref{eq:SpaceHeadway}) in 
mind, we therefore adopt the following convention from this moment on: we assume 
that, without loss of generality, for a vehicle $i$ its length $L_{i} = 1$~cell. 
This means that when the vehicle is residing in a compact jam, its space headway 
$h_{s_{i}} = 1$~cell and its space gap is consequently $g_{s_{i}} = 0$~cells 
(i.e., equivalent to the number of empty cells $d_{i}$ in front of the vehicle). 
This abstraction gives us a rigorous justification to formulate the TCA's update 
rules more intuitively using space gaps.

  \subsection{Performing measurements}

In order to characterize the behavior of a TCA model, we perform \emph{global} 
measurements on the system's lattice. These measurements are expressed as 
macroscopic quantities, defining the global density $k$, the space mean speed 
$\overline v_{s}$, and the flow $q$ as:

\begin{eqnarray}
  k               & = & \frac{N}{K},\label{eq:GlobalDensity}\\
  \overline v_{s} & = & \frac{1}{N} \sum_{i = 1}^{N}{v_{i}},\label{eq:GlobalSpaceMeanSpeed}\\
  q               & = & k \overline v_{s},\label{eq:GlobalFlow}
\end{eqnarray}

\noindent
with $N$ the number of vehicles in the system and $K$ the number of cells in the 
lattice. The above measurements are calculated every time step, and they should 
be averaged over a large measurement period $T_{\mbox{sim}}$ in order to allow 
the system to settle into an equilibrium.\\

Correlation plots of these aggregate quantities lead to time-inde\-pen\-dent 
graphs conventionally called `fundamental diagrams' \footnote{The concept of a 
`fundamental diagram' initially stems from the seminal paper of Lighthill and 
Whitham \cite{LIGHTHILL:55}, in which they assumed that an equilibrium relation 
between the flow $q$ and the concentration $k$ exists (note that the term 
`density' was not used then). This formed the basis for their derivation of a 
kinematic wave theory for traffic flows based on fluid dynamics.} (e.g., 
Fig.~\ref{fig:VDRTCANormalFundamentalDiagram}). An important remark is that is 
that these diagrams are actually \emph{only valid under stationary traffic 
conditions} and for homogeneous traffic, which is almost never the case. As 
such, real-life traffic does not `lie' on this curve because it represents the 
`average behavior' of the vehicles in a traffic stream. And although we should 
more correctly refer to our measurements as points in a certain \emph{phase 
space} (e.g., the ($k$,$q$) phase space), we will still use the terminology of 
`fundamental diagram' in the remainder of this paper when we are in fact 
referring to this phase space.

Note that we also can calculate these macroscopic quantities as \emph{local} 
averages (i.e., local in time and space), but these lead to the same results as 
for large systems and measurements periods \cite{NAGEL:95c}.\\

\clearpage

The previous global macroscopic measurements (density, average speed, and flow) 
from equations (\ref{eq:GlobalDensity}), (\ref{eq:GlobalSpaceMeanSpeed}), and 
(\ref{eq:GlobalFlow}), can be related to the microscopic equations 
(\ref{eq:SpaceHeadway}) and (\ref{eq:TimeHeadway}) as follows:

\begin{eqnarray}
  k & \propto & \overline h_{s}^{-1}, \label{eq:SpaceHeadwayAndDensity}\\
  q & \propto & \overline h_{t}^{-1}, \label{eq:TimeHeadwayAndFlow}
\end{eqnarray}

\noindent with $\overline h_{s}$ and $\overline h_{t}$ the \emph{average} space 
and time headway respectively. Note that with respect to the time gaps and time 
headways, we will work in the remainder of this paper with the \emph{median} 
instead of the arithmetic mean because the former gives more robust results when 
$h_{t_{i}},g_{t_{i}} \rightarrow +\infty$ for a vehicle $i$.\\

All the fundamental diagrams in this paper, were calculated using systems of 
$10^{3}$ cells. The first $10^{3}$~s of each simulation were discarded in order 
to let initial transients die out; the system was then updated for 
$T_{\mbox{sim}} = 10^{4}$~s.

For a deeper insight into the behavior of the space mean speed $\overline 
v_{s}$, the average space gap $\overline g_{s}$, and the median time gap 
$\overline g_{t}$, detailed histograms showing their distributions are provided. 
These are interesting because in the existing literature (e.g., 
\cite{CHOWDHURY:98,SCHADSCHNEIDER:00,HELBING:01}) these distributions are only 
considered at several distinct global densities, whereas we show them for 
\emph{all} densities. Each of our histograms is constructed by varying the 
global density $k$ between 0.0 and 1.0, calculating the average speed, the 
average space gap and the median time gap for each simulation run. A simulation 
run consists of $5 \times 10^{4}$~s (with a transient period of 500~s) on 
systems of 300~cells, varying the density in 150 steps. Note that a larger size 
of the system's lattice, has no significant effects on the results, except for 
an increase of the variance.\\

All the experiments were carried out with our Java software \emph{"Traffic 
Cellular Automata"}, which can be found at \texttt{http://smtca.dyns.cx} 
\cite{MAERIVOET:04d}.

\section{Velocity dependent randomization}
\label{sec:VDR}

In this section, the rule set of the VDR-TCA model is explained, followed by a 
an overview of the model's tempo-spatial behavior and its related macroscopic 
quantities (i.e., the fundamental diagrams and the distributions of the speeds 
and the space and time gaps) under normal conditions.

  \subsection{The VDR-TCA's rule set}
  \label{subsec:VDRTCARuleSet}

As indicated before, we focus our research on the VDR-TCA model for the 
implementation of the car-following behavior. The following equations (based on 
\cite{BARLOVIC:98}) form its rule set; the rules are applied consecutively to 
all vehicles in parallel (i.e., synchronous updates):

\begin{description}
  \item[\textbf{R1}:] \emph{determine stochastic noise}
    \begin{equation}
    \label{eq:VDRTCA-R1}
      \left \lbrace
        \begin{array}{lcl}
          v_{i}(t - 1) = 0 & \Longrightarrow & p' \leftarrow p_{0}, \\
          v_{i}(t - 1) > 0 & \Longrightarrow & p' \leftarrow p, \\
        \end{array}
      \right.
    \end{equation}

  \item[\textbf{R2}:] \emph{acceleration and braking}
    \begin{equation}
    \label{eq:VDRTCA-R2}
      v_{i}(t) \leftarrow \min \lbrace v_{i}(t - 1) + 1,g_{s_{i}}(t - 1),v_{\mbox{max}} \rbrace,
    \end{equation}

  \item[\textbf{R3}:] \emph{randomization}
    \begin{equation}
    \label{eq:VDRTCA-R3}
      \xi_{i}(t) < p' \Longrightarrow v_{i}(t) \leftarrow \max \lbrace 0,v_{i}(t) - 1 \rbrace,
    \end{equation}

  \item[\textbf{R4}:] \emph{vehicle movement}
    \begin{equation}
    \label{eq:VDRTCA-R4}
      x_{i}(t) \leftarrow x_{i}(t - 1) + v_{i}(t).
    \end{equation}
\end{description}

In the above equations, $v_{i}(t)$ is the speed of vehicle $i$ at time $t$ 
(i.e., in the current \emph{updated} configuration of the system), 
$v_{\mbox{max}}$ is the maximum allowed speed, $g_{s_{i}}$ denotes the space gap 
of vehicle $i$ and $x_{i} \in \lbrace 1,\ldots,K \rbrace$ an integer number 
denoting its position in the lattice. In the third rule, equation 
(\ref{eq:VDRTCA-R3}), $\xi_{i}(t) \in [0,1[$ denotes a uniform random number 
(specifically drawn for vehicle $i$ at time $t$) and $p'$ is the stochastic 
noise parameter, \emph{dependent on the vehicle's speed} ($p_{0}$ is called the 
slow-to-start probability and $p$ the slowdown probability, with $p_{0},p \in 
[0,1]$).

In a nutshell, rule \emph{R}1, equation (\ref{eq:VDRTCA-R1}), determines the 
correct velocity dependent randomization. Note that we only consider the case 
with two different noise parameters (i.e., $p_{0}$ and $p$), ignoring the more 
general case where we have a noise parameter for each possible speed (i.e., 
$p_{0}$, \ldots, $p_{v_{\mbox{max}}}$). Rule \emph{R}2, equation 
(\ref{eq:VDRTCA-R2}), states that a vehicle tries to increase its speed at each 
time step, as long as it hasn't reached its maximal speed and it has enough 
space headway. It also states that when a vehicle hasn't enough space headway, 
it \emph{abruptly} adapts its speed in order to prevent a collision with the 
leading vehicle (the model thus exhibits an infinite braking capability). The 
randomization parameter determined in equation (\ref{eq:VDRTCA-R1}), is now used 
in rule \emph{R}3, equation (\ref{eq:VDRTCA-R3}), to introduce a stochastic 
component in the system: a vehicle will randomly slow down with probability 
$p'$. The last rule \emph{R}4, equation (\ref{eq:VDRTCA-R4}), isn't actually a 
`real' rule; it just allows the vehicles to advance in the system.

  \subsection{Normal behavioral characteristics}
  \label{subsec:NormalBehavioralCharacteristics}

\begin{figure}[!ht]
  \centering
  \psfrag{580s}[][]{\footnotesize{Time (580~s)}}
  \psfrag{300cells}[][]{\footnotesize{Space (300~cells)}}
  \includegraphics[width=8.6cm]{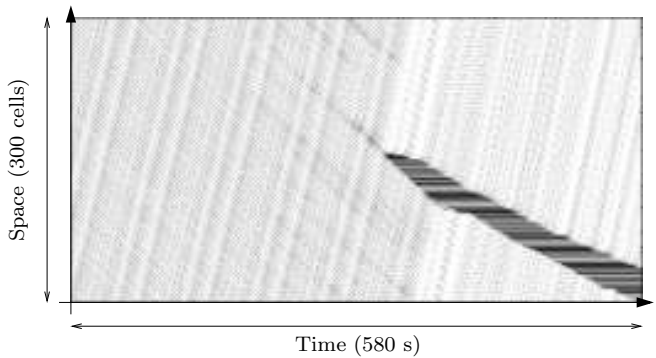}
  \caption{
    A typical time-space diagram of the VDR-TCA model. The shown lattice 
    contains 300 cells (corresponding to 2.25~km), the visible time horizon is 
    580 seconds. The slow-to-start probability $p_{0}$ was set to 0.5, the 
    slowdown probability $p$ to 0.01, and the global density equal to the 
    system's critical density. We can see a phase separation taking place, 
    forming a persistent compact jam surrounded by free-flowing traffic. The 
    significant decrease of the density in the regions outside the jam results 
    from the jam's reduced outflow.
  }
  \label{fig:VDRTCATypicalTXDiagram}
\end{figure}

Depending on their speed, vehicles are subject to different randomizations: 
typical metastable behavior results when $p_{0} \gg p$, meaning that stopped 
vehicles have to wait longer before they can continue their journey (i.e, they 
are `slow-to-start'). This has the effect of a reduced outflow from a jam, so 
that, in a closed system, this leads to an equilibrium and the formation of a 
\emph{compact} jam. For such a typical situation (e.g., $p_{0} = 0.5$ and $p = 
0.01$), the tempo-spatial evolution is depicted in 
Fig.~\ref{fig:VDRTCATypicalTXDiagram}; each vehicle is represented by a single 
colored dot; as time advances, vehicles move to the upper right corner, whereas 
congestion waves move to the lower right corner, i.e., backwards in space (note 
that the orientation of the axes is different than the one 
Fig.~\ref{fig:TCABasics}). We can see an initially homogeneous traffic pattern 
(one \emph{metastable} phase) breaking down and kicking the system into a 
\emph{phase separated state} (two phases: free-flowing traffic and a compact 
jam).

\begin{figure}[!ht]
  \centering
  \psfrag{Global density k}[][]{\footnotesize{Global density $k$}}
  \psfrag{Flow q [vehicles/hour]}[][]{\footnotesize{Flow $q$ [vehicles/h]}}
  \psfrag{Homogeneous state}[l][l]{\footnotesize{Homogeneous state}}
  \psfrag{Superjam}[l][l]{\footnotesize{Superjam}}
  \includegraphics[width=8.6cm]{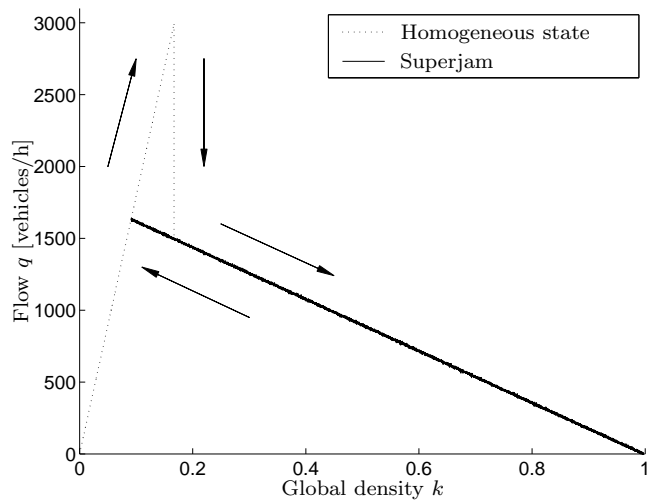}
  \caption{
    Typical ($k$,$q$) fundamental diagram of the VDR-TCA (with the slow-to-start 
    probability $p_{0}$ set to 0.5 and the slowdown probability $p$ set to 
    0.01). The  typical `reversed $\lambda$' shape is obtained by going from the 
    free-flowing to the congested traffic regime and vice versa, resulting in 
    hysteretic behavior (i.e., the directions of the arrows); in the first case, 
    traffic encounters a capacity drop, whereas in the second case, the recovery 
    results in a lower value for the flow.
  }
  \label{fig:VDRTCANormalFundamentalDiagram}
\end{figure}

Studying the ($k$,$q$) fundamental diagram in 
Fig.~\ref{fig:VDRTCANormalFundamentalDiagram}, gives us another view of this 
phase transition. The free-flow branch was obtained by initially distributing 
the vehicles homogeneously over the system's lattice, whereas they were placed 
in a compact superjam (i.e., all vehicles are `bunched up' behind each other) in 
order to obtain the congested branch. We can see a `capacity drop' taking place 
at the critical density, where traffic in its vicinity behaves in a metastable 
manner. This metastability is characterized by the fact that sufficiently large 
disturbances of the fragile equilibrium can cause the flow to undergo a sudden 
decrease, corresponding to a first-order phase transition in statistical 
physics. The state of very high flow is then destroyed and the system settles 
into a phase separated state with a large megajam and a free-flow zone 
\cite{BARLOVIC:98,CHOWDHURY:00}. The large jam will persist as long as the 
density is not significantly lowered, meaning that recovery of traffic from 
congestion thus shows a hysteresis phenomenon \cite{BARLOVIC:99}.\\

Considering the non-stochastic CA-184 (i.e., Wolfram's rule 184 with 
$v_{\mbox{max}} > 1$ \cite{WOLFRAM:02}), we can calculate the critical density 
(where the transition between the free-flowing and congested phases occurs) as 
\cite{SCHADSCHNEIDER:02}:

\begin{equation}
\label{eq:CA184CriticalDensity}
  k_{c} = \frac{1}{L + v_{\mbox{max}}}.
\end{equation}

\noindent
Setting $v_{\mbox{max}} = 5$~cells/s (this equals $5 \times 7.5 \times 3.6 = 
135$~km/h), we correspondingly get $k_{c} = (1 + 5)^{-1} = 0.1\overline6$. Note 
that this number is dimensionless because $k_{c}$ represents the fraction of the 
maximum density, i.e., the jam density $k_{j} = (\Delta X)^{-1} = 
133.\overline3$~vehicles/km. Converting the value of the critical density to 
real world measurement units, we obtain $k_{c} = 0.1\overline6 \times 
133.\overline3 = 22.\overline 2$~vehicles/km. The associated capacity flow 
$q_{\mbox{cap}} = k_{c} \cdot \overline v_{\mbox{max}} = 135 \times 
22.\overline2 = 3000$~vehicles/h. Note that this number is an overestimate for 
our system, as an exact formula expressing the capacity flow $q_{\mbox{cap}}$ in 
function of the maximum speed $v_{\mbox{max}}$, the slowdown probability $p$, 
and the lattice size $K$, has not yet been determined \cite{MAERIVOET:03g}. If 
we perform a linear regression on the congested branch of 
Fig.~\ref{fig:VDRTCANormalFundamentalDiagram}, we get the speed of the backward 
propagating congestion waves, which is $\approx -13.5$~km/h, close to the 
empirical value of -15~km/h found in real-life traffic \cite{HELBING:01}.\\

\begin{figure*}[!ht]
  \centering
  \psfrag{Global density k}[][]{\footnotesize{Global density $k$}}
  \psfrag{Speed v [cells/s]}[][]{\footnotesize{Speed $v$ [cells/s]}}
  \psfrag{(A)}[][]{\footnotesize{($A$)}}
  \psfrag{(B)}[][]{\footnotesize{($B$)}}
  \psfrag{Histogram class [speed v in cells/s]}[][]{}
  \psfrag{Histogram class probability}[][]{\footnotesize{Class probability}}
  \begin{tabular}{cc}
    \begin{tabular}{c}
      \includegraphics[width=8.2cm]{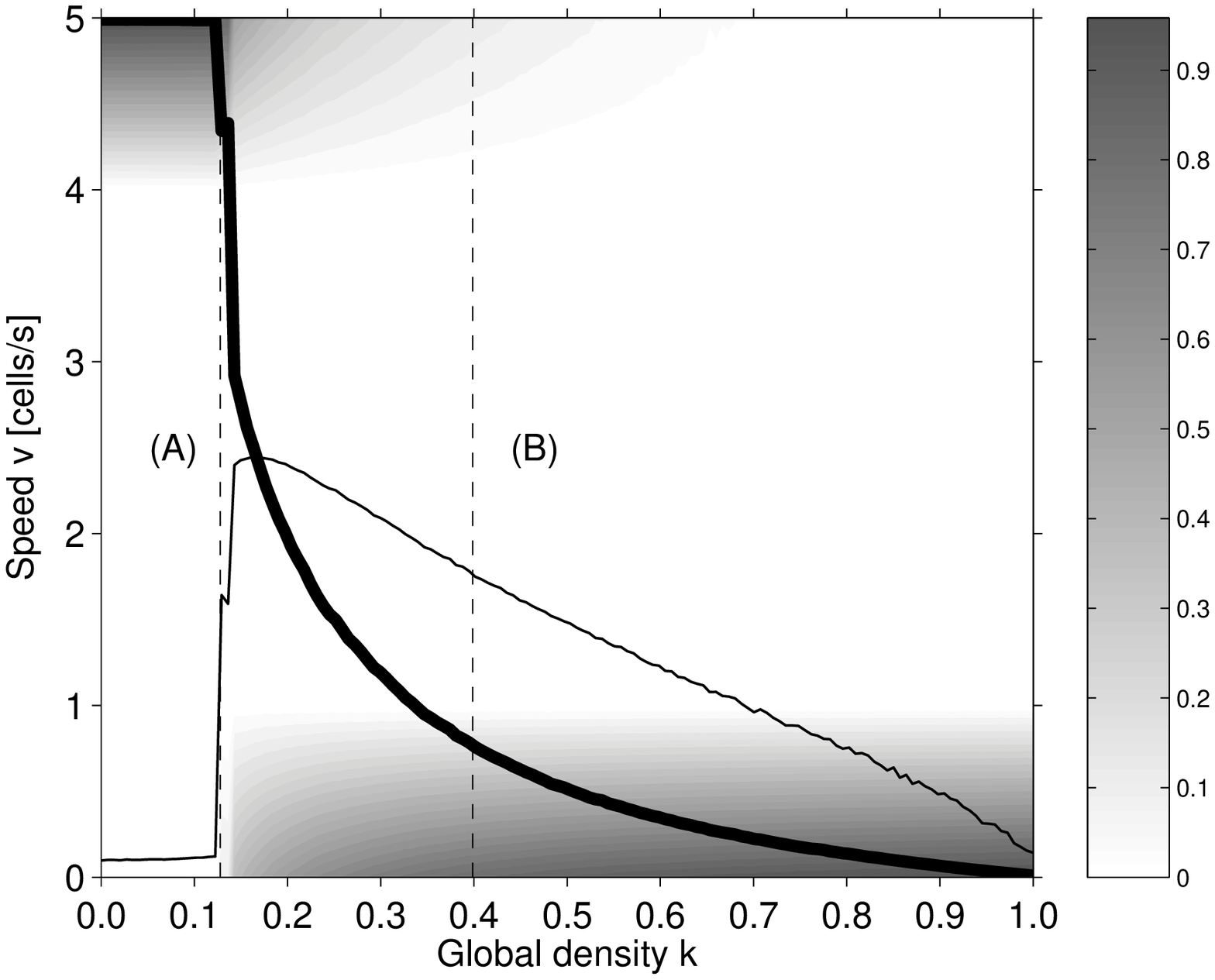}
    \end{tabular}
    &
    \begin{tabular}{c}
      \includegraphics[width=8.2cm,height=3cm]{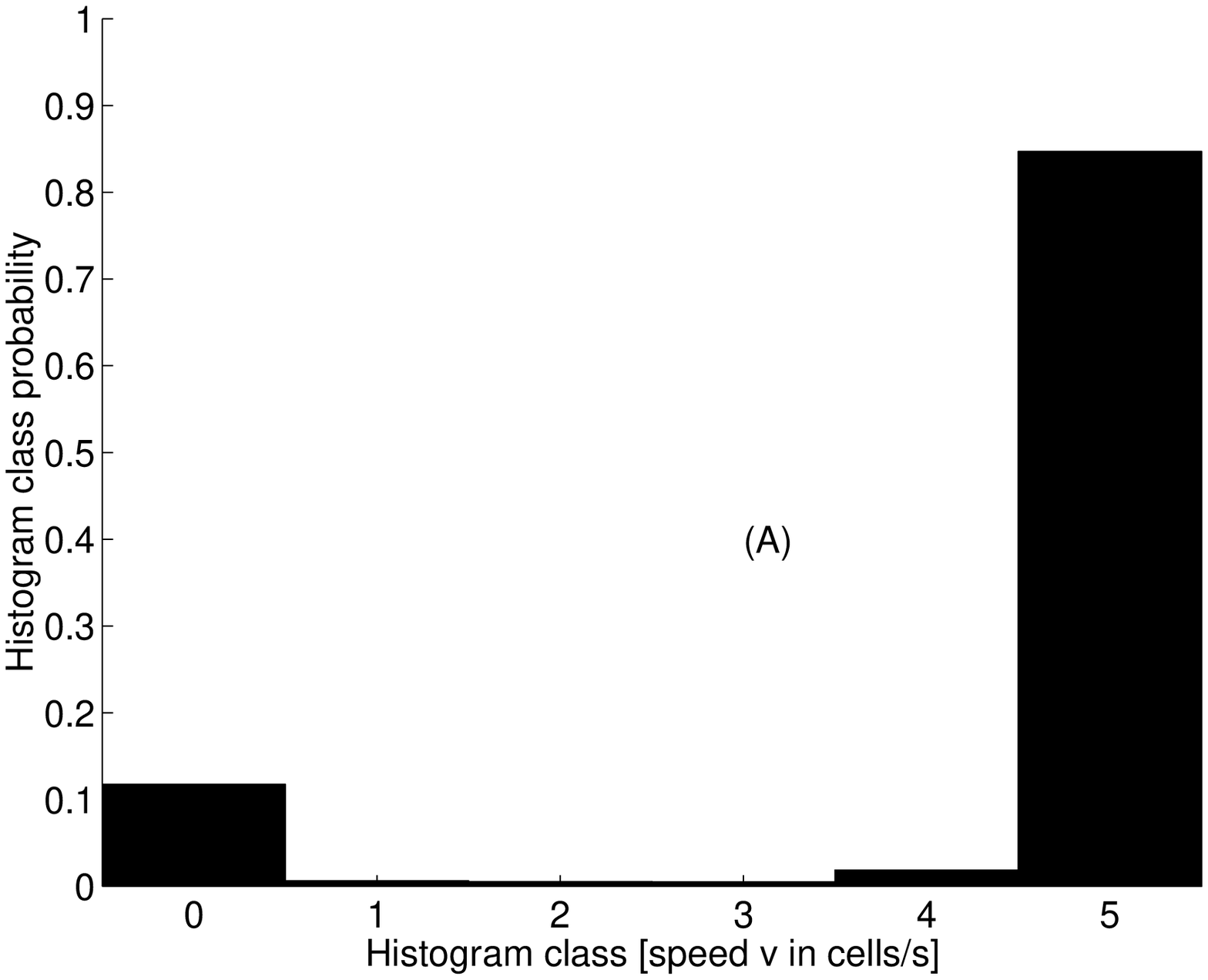}\\
      \includegraphics[width=8.2cm,height=3cm]{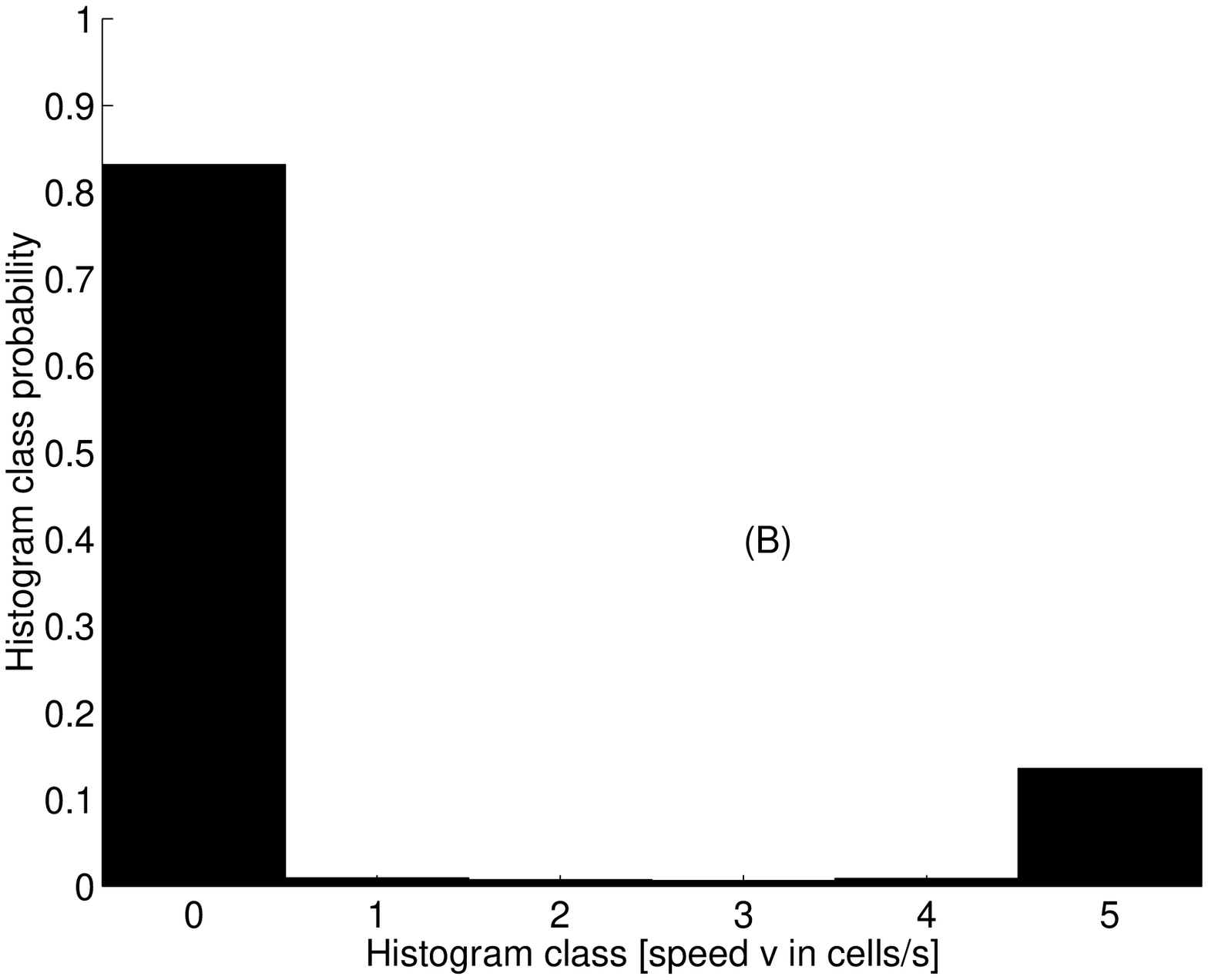}\\
      \footnotesize{Histogram class [speed $v$ in cells/s]}
    \end{tabular}
  \end{tabular}
  \caption{
    The distribution of the vehicles' speeds $v$, as a function of the global 
    density $k$ in the VDR-TCA (with $p_{0} = 0.5$ and $p = 0.01$). In the 
    contourplot to the left, the thick solid line denotes the space mean speed, 
    whereas the thin solid line shows its standard deviation. The grey regions 
    denote the probability densities. The histograms ($A$) and ($B$) to the 
    right, show two cross sections made in the left contourplot at $k = 0.1325$ 
    and $k = 0.4000$ respectively: for example, in ($A$), the high concentration 
    of probability mass at the histogram class $v = $~5 cells/s corresponds to 
    the dark region in the upper left corner of the contourplot.
  }
  \label{fig:VDRTANormalConditionsSpeedHistogram}
\end{figure*}

If we look at the distribution of the vehicles' speeds, we get the histogram in 
Fig.~\ref{fig:VDRTANormalConditionsSpeedHistogram}. Here we can clearly see the 
distinction between the free-flowing and the congested phase: the space mean 
speed remains constant at a high value, then encounters a sharp transition 
(i.e., the capacity drop), resulting in a steady declination as the global 
density increases. Note that as the critical density is encountered, the 
standard deviation jumps steeply; this means that vehicles' speeds fluctuate 
wildly at the transition point (because they are entering and exiting the 
congestion waves). Once the compact jam is formed, the dominating speed quickly 
becomes zero (because vehicles are standing still inside the jam). Although most 
of the weight is attributed to this zero-speed, there is a non-negligible 
maximum speed present for intermediate densities. If the global density is 
increased further towards the jam density, this maximum speed disappears and the 
system settles into a state in which all vehicles either have speed zero or one 
(i.e., systemwide stop-and-go traffic).\\

\begin{figure}[!ht]
  \centering
  \psfrag{Global density k}[][]{\footnotesize{Global density $k$}}
  \psfrag{Space gap gs [cells]}[][]{\footnotesize{Space gap $g_{s}$ [cells]}}
  \psfrag{>= 10}[][]{\footnotesize{$\geq$~10}}
  \includegraphics[width=8.6cm]{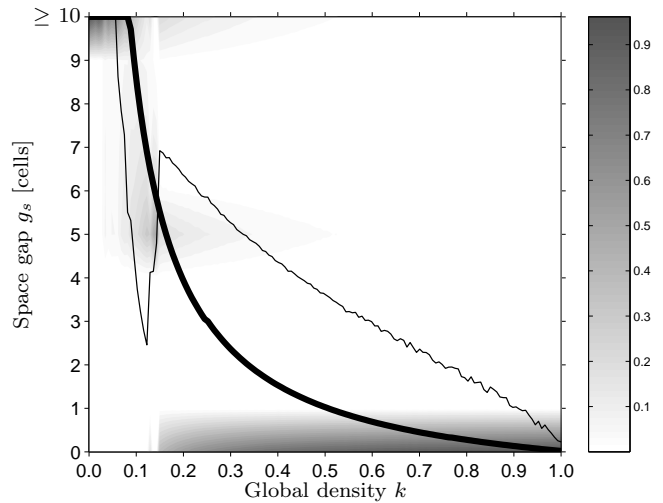}
  \caption{
    The distribution of the vehicles' space gaps $g_{s}$, as a function of the 
    global density $k$ in the VDR-TCA (with $p_{0} = 0.5$ and $p = 0.01$). The 
    thick solid line denotes the average of all the vehicles' space gaps, 
    whereas the thin solid line shows its standard deviation. The grey regions 
    denote the probability densities.
  }
  \label{fig:VDRTANormalConditionsSpaceGapHistogram}
\end{figure}

Considering the distribution of the vehicles' space gaps, we get 
Fig.~\ref{fig:VDRTANormalConditionsSpaceGapHistogram}; because of their tight 
coupling in the VDR-TCA's rule set, the courses of both the space mean speed and 
the average space gaps are similar. Although the space gaps are rather large for 
low densities, at the critical density $k_{c}$ they leave a small cluster around 
an optimal value. This value can be calculated using equations 
(\ref{eq:SpaceHeadway}), (\ref{eq:SpaceHeadwayAndDensity}), and 
(\ref{eq:CA184CriticalDensity}) (noting that $\overline L = L = 1$ because we 
are dealing with homogeneous traffic):

\begin{equation}
  \overline g_{s} = \overline h_{s} - \overline L = \frac{1}{k_{c}} - L,
\end{equation}

\noindent
resulting in $\lceil \overline g_{s} \rceil = \lceil 0.1669^{-1} - 1 \rceil = 
\lceil 4.99 \rceil = 5$~cells. This corresponds to the necessary minimal space 
gap in order to travel at the maximum speed, avoiding a collision with the 
leader. An important observation is that no recorded space gaps exist between 
this cluster of five cells and the space gaps of zero cells inside the compact 
jam. This means that there is a \emph{distinct phase separation} taking place 
once beyond the critical density: vehicles are either completely in the 
free-flowing regime, or they are in the compact jam.\\

\begin{figure}[!ht]
  \centering
  \psfrag{Global density k}[][]{\footnotesize{Global density $k$}}
  \psfrag{Time gap gt [seconds]}[][]{\footnotesize{Time gap $g_{t}$ [seconds]}}
  \psfrag{+INF}[][]{\footnotesize{$+\infty$}}
  \includegraphics[width=8.6cm]{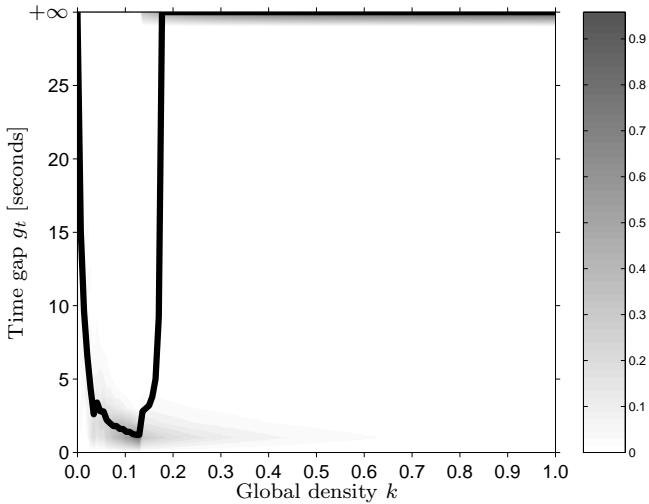}
  \caption{
    The distribution of the vehicles' time gaps $g_{t}$, as a function of the 
    global density $k$ in the VDR-TCA (with $p_{0} = 0.5$ and $p = 0.01$). The 
    thick solid line denotes the \emph{median} of all the vehicles' time gaps. 
    The grey regions denote the probability densities.
  }
  \label{fig:VDRTANormalConditionsTimeGapHistogram}
\end{figure}

Figure~\ref{fig:VDRTANormalConditionsTimeGapHistogram} shows the distribution of 
the vehicles' time gaps: when traffic is in the free-flowing regime, time gaps 
are high, \emph{but finite}. As the critical density is approached, the median 
time gap first decreases (because the vehicles' speeds remain the same but their 
space gaps decrease). Once beyond the critical density, it increases 
\emph{towards infinity} because vehicles come to a full stop inside the compact 
jam. Just as with the space gaps, we can also observe a small cluster around an 
optimal value. This optimal time gap, is the time needed to travel the distance 
formed by the optimal space gap, at the maximum speed. This means that 
$\overline g_{t} = \overline g_{s} \div v_{\mbox{max}} = 1$~s. Because the 
previous train of thought is only exactly valid for the non-stochastic CA-184 
and the VDR-TCA incorporates stochastic noise, the \emph{median} optimal time 
gap lies somewhat above the previously calculated value ($\overline g_{t} 
\approx 1.2$~s).

\section{More complex system dynamics}
\label{section:Results}

Most of the previous research dealt with the study of the behavioral 
characteristics of the VDR-TCA model operating under normal conditions. We now 
turn our attention to the specific case in which the model's parameters take on 
extreme values $p_{0} \ll p$, more specifically considering the limiting case 
where $p_{0} = 0.0$ and $p = 1.0$.\\

We will first look at the change in tempo-spatial behavior when $p$ is increased 
towards 1.0, at which point a peculiar behavior is established in the system. We 
study the qualitative effects that the VDR-TCA's rule set has on individual 
vehicles, and discuss shortly the prevailing initial conditions. This is 
followed by a quantitative analysis using the ($k$,$q$) and ($k$,$\overline 
v_{s}$) fundamental diagrams. The former diagram leads us to the discovery of 
four distinct phases, having a non-concave region with forward propagating 
density waves. More elaborate explanations are given based on the histograms of 
the space mean speed, the average space gap, and the median time gap. This is 
followed by a detailed analysis of the observations of the tempo-spatial 
behavior of the system in each of the four different phases (i.e., traffic 
regimes). The section concludes with a short discussion on the effects of 
different maximal speeds, after which we adopt the use of an order parameter 
that is able to track the phase transitions between the different traffic 
regimes.

  \subsection{Increasing the stochastic noise $p$}

\begin{figure}[!ht]
  \centering
  \psfrag{580s}[][]{\footnotesize{Time (580~s) with $p \in [0 \rightarrow 1]$}}
  \psfrag{300cells}[][]{\footnotesize{Space (300~cells)}}
  \includegraphics[width=8.6cm]{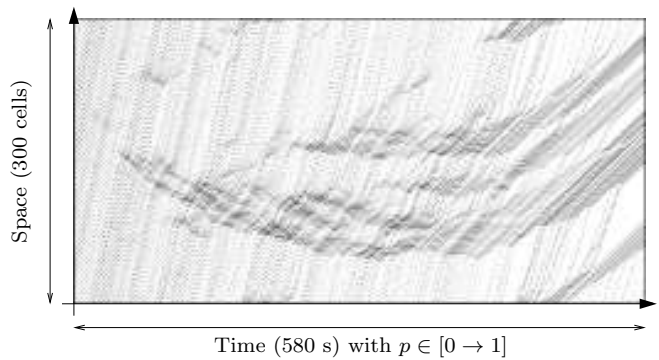}
  \caption{
    A time-space diagram showing a system with a lattice of 300 cells 
    (corresponding to 2.25~km); the visible time horizon is 580 seconds. As time 
    advances, the slowdown probability $p$ varies between 0.0 and 1.0 (the 
    slow-to-start probability $p_{0}$ was fixed at 0.0). The system's global 
    density $k$ is 0.1667.
  }
  \label{fig:TXDiagramVDRTCA-SP0-P01}
\end{figure}

Let us first consider the case in which $p_{0} = 0.0$ and where we vary $p$ 
between 0.0 and 1.0. Figure~\ref{fig:TXDiagramVDRTCA-SP0-P01} shows a time-space 
diagram where we simulated a system consisting of 300~cells. As time advances 
(over a period of 580~s), the slowdown probability $p$ is steadily increased 
from 0.0 to 1.0 (the global density $k$ was set to 0.1667 which is slightly 
below the critical density for a system with stochastic noise).

When $p$ is very low, vehicles can keep driving in the free-flowing regime. As 
$p$ is increased, small unstable jams occur. Increasing $p$ even further, leads 
to even more pronounced jams. An important observation is that the propagation 
speed of these backward moving congestion waves \emph{increases}; when $p$ 
reaches approximately 0.5, their speed equals 0, meaning that jams stay fixed at 
a position. Note that as these jams are unstable, they can be created or 
dissolved at arbitrary locations. Finally, as $p$ tends to 1.0, we can see the 
emergence of \emph{forward propagating} congestion waves; `density' is now being 
carried in the direction of the traffic flow.  These forward propagating waves 
form `moving blockades' that trap vehicles, which in turn leads to tightly 
packed clusters of vehicles that move steadily (but at a much slower pace than 
in the free-flowing regime). Note that this `clustering' behavior is different 
from platooning, which typically occurs when vehicles are driving close to each 
other at relatively high speeds \cite{LARRAGA:04}. In our case, the clusters of 
vehicles advance more slowly.

  \subsection{Qualitative effects of the rule set}
  \label{subsec:QualitativeEffectsRuleSet}

From now on, we only consider the limiting case where $p_{0} = 0.0$ and $p = 
1.0$. Let us now study the influence of the VDR-TCA's rule set on an individual 
vehicle $i$. Assuming $v_{i} \in \lbrace 0, \ldots, v_{\mbox{max}}\rbrace$, the 
rules described in section \ref{subsec:VDRTCARuleSet} lead to the following four 
general cases:

\begin{itemize}

  \item Case (1) with $v_{i}(t - 1) = 0$ and $g_{s_{i}}(t - 1) = 0$

In this case, the vehicle is residing \emph{inside} a jam and rule \emph{R}2 
(acceleration and braking) plays a dominant part: the vehicle's speed 
$v_{i}(t)$ remains 0.\\

  \item Case (2) with $v_{i}(t - 1) = 0$ and $g_{s_{i}}(t - 1) > 0$

This situation may arise when, for example, a vehicle is at a jam's front. 
According to rule \emph{R}1, the stochastic noise parameter $p'$ becomes 0.0. 
Rule \emph{R}2 then results in an updated speed $v_{i}(t) = \min \lbrace 
1,g_{s_{i}}(t - 1) \rbrace = 1$. Because there is no stochastic noise present in 
this case, rule \emph{R}3 does not apply and the vehicle always advances one 
cell.\\

  \item Case (3) with $v_{i}(t - 1) > 0$ and $g_{s_{i}}(t - 1) = 0$

Because there is no space in front of the vehicle, it has to brake in order to 
avoid a collision. Rule \emph{R}2 consequently abruptly decreases the vehicle's 
speed $v_{i}(t)$ to 0, as the vehicle needs to stop.\\

  \item Case (4) with $v_{i}(t - 1) > 0$ and $g_{s_{i}}(t - 1) > 0$

This case deserves special attention, as there are now two discriminating 
possibilities:\\

  \begin{itemize}

    \item Case (4a) with $v_{i}(t - 1) < g_{s_{i}}(t - 1)$

    According to rule \emph{R}1, the stochastic noise $p'$ becomes 1.0. Because 
    the vehicle's speed is strictly less than its space gap, rule \emph{R}2 
    becomes $v_{i}(t) \leftarrow \min \lbrace v_{i}(t - 1) + 1,v_{\mbox{max}} 
    \rbrace$. Finally, rule \emph{R}3 is applied which \emph{always} reduces the 
    speed calculated in rule \emph{R}2 (constrained to 0). In order to 
    understand what is happening, consider the speeds $v_{i}(t - 1)$ and 
    $v_{i}(t)$ in the following table:

    \begin{table}[!ht]
      \centering
      \begin{tabular}{lcl}
        $v_{i}(t - 1)$       &                   & $v_{i}(t)$\\
        \hline
        $v_{\mbox{max}}$     & $\longrightarrow$ & $v_{\mbox{max}}$ - 1\\
        $v_{\mbox{max}}$ - 1 & $\longrightarrow$ & $v_{\mbox{max}}$ - 1\\
        $v_{\mbox{max}}$ - 2 & $\longrightarrow$ & $v_{\mbox{max}}$ - 2\\
        \vdots               &                   & \vdots\\
        2                    & $\longrightarrow$ & 2\\
        1                    & $\longrightarrow$ & 1\\
      \end{tabular}
    \end{table}

    We can clearly see that the maximum speed a vehicle can travel at, is 
    constrained by $v_{\mbox{max}}$ - 1, which corresponds to $v_{\mbox{max}} - 
    p'$ \cite{SCHRECKENBERG:01}. From the table it follows that all vehicles 
    traveling at $v_{i} < v_{\mbox{max}}$ can \emph{neither accelerate nor 
    decelerate}: the vehicles' current speed is kept.\\

  \item Case (4b) with $v_{i}(t - 1) \geq g_{s_{i}}(t - 1)$

    Just as in the previous case (4a), the stochastic noise $p'$ becomes 
    1.0. Rule \emph{R}2 now changes to $v_{i}(t) \leftarrow g_{s_{i}}(t - 1)$. 
    Because rule \emph{R}3 is \emph{always} applied, this results in $v_{i}(t)$ 
    actually becoming $g_{s_{i}}(t - 1) - 1$ instead of just $g_{s_{i}}(t - 1)$. 
    So a vehicle \emph{always} slows down \emph{too much} (as opposed to solely 
    avoiding a collision with its leader).
  \end{itemize}
\end{itemize}

\noindent
\textbf{Conclusion:} considering the previously discussed four general cases, 
the most striking feature is that, according to case (4a), in a VDR-TCA 
model with $p_{0} = 0.0$ and $p = 1.0$, a moving vehicle can \emph{never} 
increase its speed, i.e., $v_{i}(t) \leq v_{i}(t - 1)$.

  \subsection{Effects of the initial conditions}
  \label{subsec:InitialConditions}

As already mentioned, in this paper we study the limiting case where $p_{0} = 
0.0$ and $p = 1.0$. This case is special, in the sense that the system's 
behavior for light densities is extremely dependent on the initial conditions.\\

\begin{figure}[!ht]
  \centering
  \psfrag{Global density k}[][]{\footnotesize{Global density $k$}}
  \psfrag{Flow q [vehicles/hour]}[][]{\footnotesize{Flow $q$ [vehicles/h]}}
  \psfrag{vMax = 5 (homogeneous)}[l][l]{\footnotesize{$v_{\mbox{max}}$ = 5 (homog.)}}
  \psfrag{vMax = 4 (homogeneous)}[l][l]{\footnotesize{$v_{\mbox{max}}$ = 4 (homog.)}}
  \psfrag{vMax = 3 (homogeneous)}[l][l]{\footnotesize{$v_{\mbox{max}}$ = 3 (homog.)}}
  \psfrag{vMax = 5 (random)}[l][l]{\footnotesize{$v_{\mbox{max}}$ = 5 (random)}}
  \includegraphics[width=8.6cm]{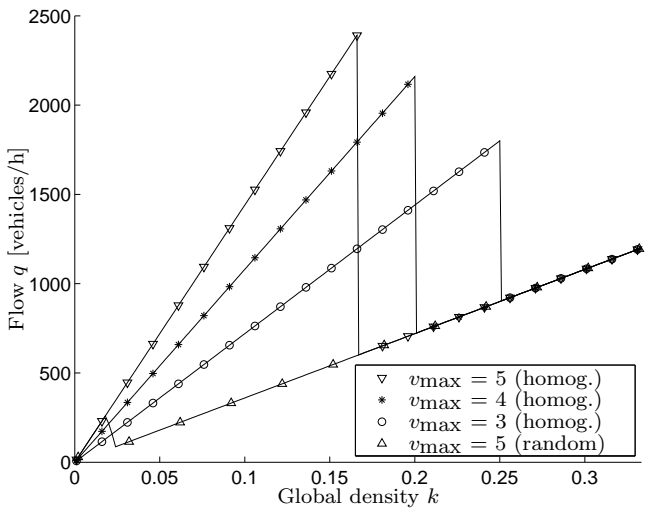}
  \caption{
    The VDR-TCA (with $p_{0} = 0.5$ and $p = 0.01$) system's behavior for light 
    densities has a strong dependence on the initial conditions. The ($k$,$q$) 
    fundamental diagrams are shown for $v_{\mbox{max}} \in \lbrace 3,4,5 
    \rbrace$ with $k \in [0,\frac{1}{3}]$. The systems have been initialized 
    using either a homogeneous or a random distribution of the vehicles.
  }
  \label{fig:VDRTCAKQFundamentalDiagramDifferentInitialConditions}
\end{figure}

If we calculate the flow $q$ associated with each global density $k$ in the 
range $[0,\frac{1}{3}]$, we obtain 
Fig.~\ref{fig:VDRTCAKQFundamentalDiagramDifferentInitialConditions}. Here we 
show three curves for $v_{\mbox{max}} \in \lbrace 3,4,5 \rbrace$ where the 
system was initialized by distributing the vehicles homogeneously over the 
system's lattice. A fourth curve is also shown, corresponding to $v_{\mbox{max}} 
= 5$ where, as opposed to the previous initialization scheme, the system was 
initialized by distributing the vehicles randomly over the system's lattice.\\

After the system has settled into an equilibrium, the resulting flow is limited 
by the fact that the maximum speed of a vehicle in the system is always equal to 
$v_{\mbox{max}} - 1$. As proven in section 
\ref{subsec:QualitativeEffectsRuleSet}, vehicles can never accelerate, which 
means that the slowest car in the system determines the maximum possible flow. 
Therefore, if the system (with $v_{\mbox{max}} = 5$~cells/s) is initialized with 
a homogeneous distribution of the vehicles, then all of them will travel at 
$v_{\mbox{max}} - 1 = 4$~cells/s (i.e., the upper branch in 
Fig.~\ref{fig:VDRTCAKQFundamentalDiagramDifferentInitialConditions}). If 
however, the system is initialized randomly, this has the effect that some 
vehicles are spaced more closely to each other. As a direct consequence of this, 
all vehicles will now travel at $v_{\mbox{max}}' - 1$~cells/s with 
$v_{\mbox{max}}'$ the speed of the slowest vehicle in the system. In the extreme 
case, a single vehicle traveling at $v_{i} = 1$~cell/s will cause all other 
vehicles to slow down to this speed, resulting in the lower branch in 
Fig.~\ref{fig:VDRTCAKQFundamentalDiagramDifferentInitialConditions}.

Note that the effects of different initial conditions (i.e., homogeneous versus 
random distributions of the vehicles) are only relevant for \emph{light} 
densities. Because the phase transitions in this paper occur at moderate to high 
densities, it doesn't matter what kind of initialization scheme is used, so from 
now on we always assume that all the vehicles are distributed randomly.

  \subsection{Quantitative analysis}

Whereas the previous sections dealt with the effects of the VDR-TCA's changed 
rule set and the role of the initial conditions, this section considers the 
effects on the ($k$,$q$) and ($k$,$\overline v_{s}$) fundamental diagrams (i.e., 
three phase transitions and a non-concave region), as well as on the histograms 
of the space mean speed, the average space gap and the median time gap.

    \subsubsection{Effects on the fundamental diagrams}

\begin{figure}[!ht]
  \centering
  \psfrag{Global density k}[][]{\footnotesize{Global density $k$}}
  \psfrag{Flow q [vehicles/hour]}[][]{\footnotesize{Flow $q$ [vehicles/h]}}
  \psfrag{p = 0.00}[][]{\footnotesize{$p = 0.00$}}
  \psfrag{p = 0.25}[][]{\footnotesize{$p = 0.25$}}
  \psfrag{p = 0.50}[][]{\footnotesize{$p = 0.50$}}
  \psfrag{p = 0.75}[][]{\footnotesize{$p = 0.75$}}
  \psfrag{p = 0.90}[][]{\footnotesize{$p = 0.90$}}
  \psfrag{p = 1.00}[][]{\footnotesize{$p = 1.00$}}
  \psfrag{(I)}[][]{\footnotesize{(\emph{I})}}
  \psfrag{(II)}[][]{\footnotesize{(\emph{II})}}
  \psfrag{(III)}[][]{\footnotesize{(\emph{III})}}
  \psfrag{(IV)}[][]{\footnotesize{(\emph{IV})}}
  \includegraphics[width=8.6cm]{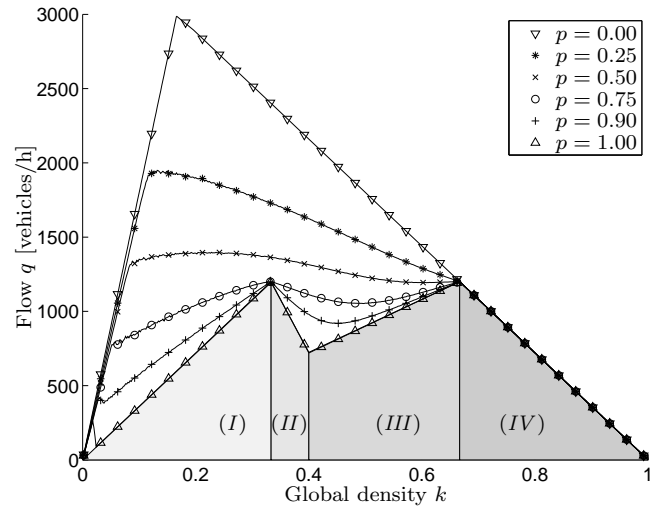}
  \caption{
    The ($k$,$q$) fundamental diagrams of the VDR-TCA with a fixed slow-to-start 
    probability $p_{0} = 0.0$ (the slowdown probability $p$ is increased from 
    0.0 to 1.0). In the limiting case where $p = 1.0$, four distinct density 
    regions (\emph{I})--(\emph{IV}) appear.
  }
  \label{fig:VDRTCAKQFundamentalDiagramsForDifferentP}
\end{figure}

Most existing ($k$,$q$) fundamental diagrams related to traffic flow models, 
show a concave course (although some (pedagogic) counter-examples such as 
\cite{AWAZU:99,GRABOLUS:01,GRAY:01,LEVEQUE:01,NISHINARI:03,ZHANG:03} exist). 
Non-concavity (i.e., convexity) of a function $f$ is defined as $\forall x_{1}, 
x_{2} \in \mbox{dom} f~|~f(\frac{1}{2} (x_{1} + x_{2})) \leq \frac{1}{2} 
f((x_{1}) + f(x_{2}))$. The property of concavity also holds true for the 
VDR-TCA model operating under normal conditions (cfr. 
Fig.~\ref{fig:VDRTCANormalFundamentalDiagram}). However, when $p_{0} \ll p$, 
this is no longer the case: in the limit where $p_{0} = 0.0$ and $p = 1.0$, the 
fundamental diagram exhibits two distinct sharp peaks. Between these peaks, 
there exists a region (\emph{II})+(\emph{III}) where the fundamental diagram has 
a convex shape, as can be seen from 
Fig.~\ref{fig:VDRTCAKQFundamentalDiagramsForDifferentP}. In region (\emph{III}), 
traffic flow has a tendency to \emph{improve} with increasing density. Note that 
we ignore the hysteretic behavior of the fundamental diagrams, as it gives no 
relevant contribution to the results of our experiments.

As the slowdown probability $p$ is increased from 0.0 to 1.0, the critical 
density -- at which the transition from the free-flowing regime occurs -- is 
shifted to lower values (note that the magnitude of the capacity drop also 
diminishes). For a global density $k = \frac{1}{6}$ (i.e., the critical density 
of the non-stochastic CA-184), we can see that the speed of the backward 
propagating congestion waves increases, just as was visible in the time-space 
diagram of Fig.~\ref{fig:TXDiagramVDRTCA-SP0-P01}. The speed of these 
characteristics is defined as $\partial q / \partial k$ (i.e., the tangent to 
the fundamental diagram), and when $p \approx 0.5$, the sign of the speed is 
reversed, leading to the earlier mentioned \emph{forward propagating} density 
waves. Furthermore, as is apparent in 
Fig.~\ref{fig:VDRTCAKQFundamentalDiagramsForDifferentP}, for high densities 
there exists a region (\emph{IV}) in which the flow $q$ is only dependent on $k$ 
and \emph{not} on the stochastic noise $p$: all the measurements coincide on 
this heavily congested branch.\\

\begin{figure}[!ht]
  \centering
  \psfrag{Global density k}[][]{\footnotesize{Global density $k$}}
  \psfrag{Speed v [cells/s]}[][]{\footnotesize{Space mean speed $\overline v_{s}$ [cells/s]}}
  \psfrag{p = 0.00}[][]{\footnotesize{$p = 0.00$}}
  \psfrag{p = 0.25}[][]{\footnotesize{$p = 0.25$}}
  \psfrag{p = 0.50}[][]{\footnotesize{$p = 0.50$}}
  \psfrag{p = 0.75}[][]{\footnotesize{$p = 0.75$}}
  \psfrag{p = 0.90}[][]{\footnotesize{$p = 0.90$}}
  \psfrag{p = 1.00}[][]{\footnotesize{$p = 1.00$}}
  \includegraphics[width=8.6cm]{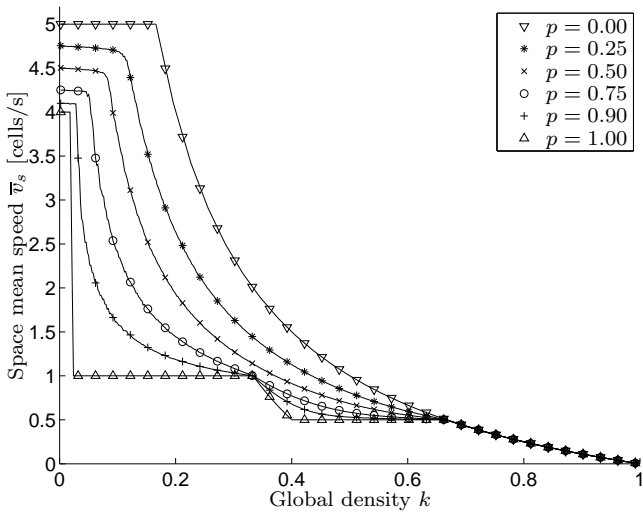}
  \caption{
    The ($k$,$\overline v_{s}$) fundamental diagrams of the VDR-TCA with a fixed 
    slow-to-start probability $p_{0} = 0.0$ (the slowdown probability $p$ is 
    increased from 0.0 to 1.0).
  }
  \label{fig:VDRTCAKVFundamentalDiagramsForDifferentP}
\end{figure}

Increasing the slowdown probability $p$ has also an effect on the space mean 
speed in the free-flowing regime. As already stated in section 
\ref{subsec:NormalBehavioralCharacteristics}, this speed is equal to 
$v_{\mbox{max}} - p$ \cite{SCHRECKENBERG:01}. The effect is more pronounced if 
we look at the ($k$,$\overline v_{s}$) fundamental diagram in 
Fig.~\ref{fig:VDRTCAKVFundamentalDiagramsForDifferentP}. Here we can see that 
the average maximum speed in the free-flowing regime shifts downwards, reaching 
a value of 4~cells/s when $p = 1.0$. From then on, there are two regions 
(\emph{I}) and (\emph{III}) where the average speed remains constant. Note that, 
to be precise, region (\emph{I}) actually contains a small capacity drop at a 
very low density, but we ignore this effect, thus treating region (\emph{I}) in 
an overall manner. Finally, note that the ($k$,$\overline v_{s}$) fundamental 
diagram remains a decreasing function as $p \rightarrow 1.0$.

    \subsubsection{Effects on the histograms}

\begin{figure}[!ht]
  \centering
  \psfrag{Global density k}[][]{\footnotesize{Global density $k$}}
  \psfrag{Speed v [cells/s]}[][]{\footnotesize{Speed $v$ [cells/s]}}
  \includegraphics[width=8.6cm]{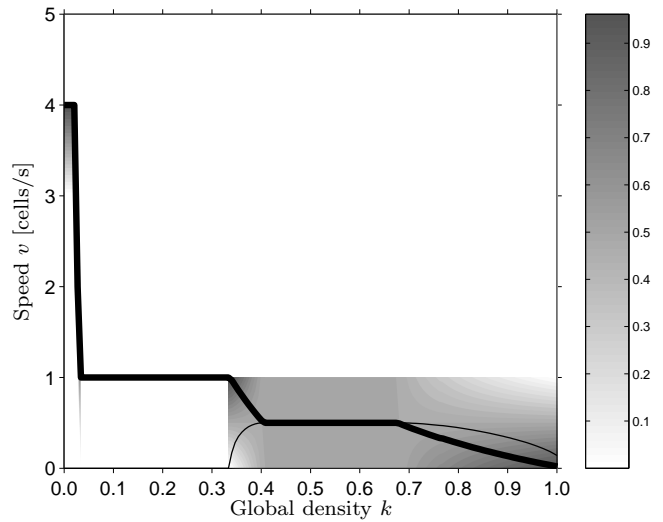}
  \caption{
    The distribution of the vehicles' speeds $v$, as a function of the global 
    density $k$ in the VDR-TCA (with $p_{0} = 0.0$ and $p = 1.0$). The thick 
    solid line denotes the space mean speed, whereas the thin solid line shows 
    its standard deviation. The grey regions denote the probability densities.
  }
  \label{fig:VDRTASpeedHistogramForHighP}
\end{figure}

The distribution of the vehicles' speeds is shown in 
Fig.~\ref{fig:VDRTASpeedHistogramForHighP}: just as in the ($k$,$\overline 
v_{s}$) fundamental diagram in 
Fig.~\ref{fig:VDRTCAKVFundamentalDiagramsForDifferentP}, we can clearly observe 
two `probability plateaus'. In the first region (\emph{I}), vehicles' speeds are 
highly concentrated in a small region around $\overline v_{s} = 1$~cell/s. As 
the global density increases in region (\emph{II}), the space mean speed 
declines until it reaches region (\emph{III}) where the second plateau is met at 
$\overline v_{s} = 0.5$~cell/s. From then on, it steadily decreases, reaching 
zero at the jam density ($k_{j} = 1.0$). Note that in region (\emph{I}), the 
standard deviation is zero, whereas it is non-zero \emph{but constant} in region 
(\emph{III}). This means that vehicles in the former region all drive \emph{at 
the same speed}; in the latter region they drive at speeds alternating between 0 
and 1~cell/s (i.e., stop-and-go traffic).

Considering the steep descending curve at the beginning of region (\emph{I}), we 
state that although vehicles can drive at $v = v_{\mbox{max}} - 1 = 4$~cell/s 
under free-flowing conditions, they nonetheless \emph{all} slow down as soon as 
at least one vehicle has a too small space gap. In other words, when $g_{s_{i}} 
\leq v_{\mbox{max}}$, case (4b) from section 
\ref{subsec:QualitativeEffectsRuleSet} applies and the vehicle slows down. This 
leads to a chain reaction of vehicles slowing down, because vehicles can never 
accelerate, as was pointed out in section \ref{subsec:InitialConditions}.\\

\begin{figure}[!ht]
  \centering
  \psfrag{Global density k}[][]{\footnotesize{Global density $k$}}
  \psfrag{Space gap gs [cells]}[][]{\footnotesize{Space gap $g_{s}$ [cells]}}
  \psfrag{>= 10}[][]{\footnotesize{$\geq$~10}}
  \includegraphics[width=8.6cm]{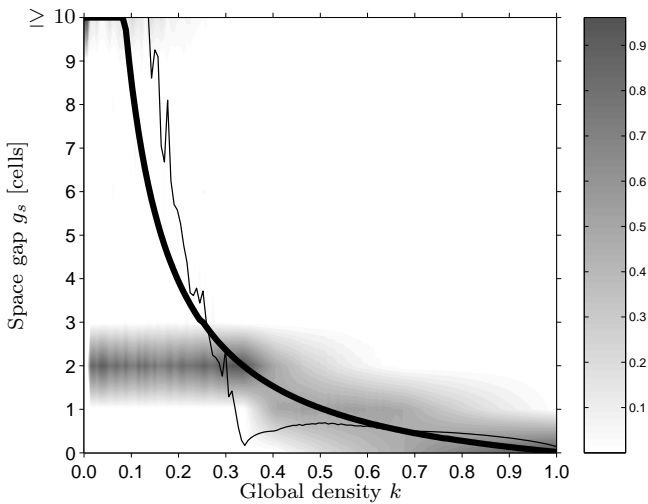}
  \caption{
    The distribution of the vehicles' space gaps $g_{s}$, as a function of the 
    global density $k$ in the VDR-TCA (with $p_{0} = 0.0$ and $p = 1.0$). The 
    thick solid line denotes the average of all the vehicles' space gaps, 
    whereas the thin solid line shows its standard deviation. The grey regions 
    denote the probability densities.
  }
  \label{fig:VDRTASpaceGapHistogramForHighP}
\end{figure}

Under normal operating conditions (i.e., $p_{0} \gg p$), a vehicle's average 
speed and average space gap show a high correlation in the congested density 
region beyond the critical density. This can be seen from the similarity between 
the histogram curves in Figs.~\ref{fig:VDRTANormalConditionsSpeedHistogram} and 
\ref{fig:VDRTANormalConditionsSpaceGapHistogram}. In high contrast with this, is 
the distribution of the vehicles' space gaps as in 
Fig.~\ref{fig:VDRTASpaceGapHistogramForHighP}, which shows a different scenario. 
More or less similar to the vehicles' speeds, we can observe the formation of 
\emph{three} plateaus of constant space gaps in certain density regions. These 
plateaus are located at 2, 1 and 0~cells for density regions (\emph{I}), 
(\emph{III}), and (\emph{IV}) respectively. Note that the average space gaps 
themselves are \emph{not} constant in these regions, as opposed to the space 
mean speed. This is due to the fact that vehicles encounter waves of stop-and-go 
traffic, whereby the frequency of these waves increases as the global density is 
augmented.

Another observation that we can make, is that the standard deviation goes to 
zero at the transition point between density regions (\emph{I}) and (\emph{II}). 
This means that, as expected, the traffic flow at this point consists of 
completely homogeneous traffic, in which all the vehicles drive with the same 
space gap $g_{s} = 2$~cells (and as already mentioned, with the same speed $v = 
1$~cell/s).\\

\begin{figure}[!ht]
  \centering
  \psfrag{Global density k}[][]{\footnotesize{Global density $k$}}
  \psfrag{Time gap gt [seconds]}[][]{\footnotesize{Time gap $g_{t}$ [seconds]}}
  \psfrag{+INF}[][]{\footnotesize{$+\infty$}}
  \includegraphics[width=8.6cm]{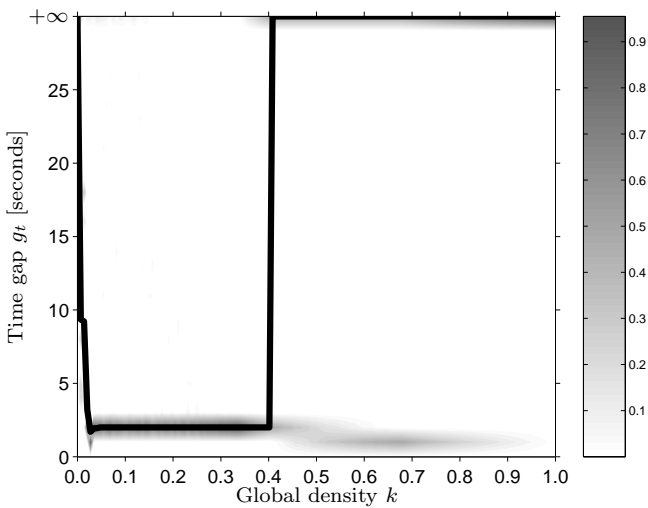}
  \caption{
    The distribution of the vehicles' time gaps $g_{t}$, as a function of the 
    global density $k$ in the VDR-TCA (with $p_{0} = 0.0$ and $p = 1.0$). The 
    thick solid line denotes the \emph{median} of all the vehicles' time gaps. 
    The grey regions denote the probability densities.
  }
  \label{fig:VDRTATimeGapHistogramForHighP}
\end{figure}

Just as with the previous histograms, there also appear to be plateaus of 
concentrated probability mass in the distribution of the vehicles' time gaps in 
Fig.~\ref{fig:VDRTATimeGapHistogramForHighP}. As opposed to the standard 
behavior in Fig.~\ref{fig:VDRTANormalConditionsTimeGapHistogram}, the 
concentration in the first region (\emph{I}) is more elongated and more or less 
completely flat. This is expected because the majority of the space mean speeds 
and the average space gaps remain constant in this region. Furthermore, once the 
first phase transition occurs, the median time gap $\overline g_{t} \rightarrow 
+\infty$ as the space mean speeds and average space gaps tend to zero. This is 
expressed as the existence of a non-neglibible cluster of probability mass at 
the top of the histogram in Fig.~\ref{fig:VDRTATimeGapHistogramForHighP}; the 
concentration is formed by vehicles that are encountering the earlier mentioned 
stop-and-go waves, resulting in the fact that their time gaps periodically 
become infinity.

  \subsection{Typical tempo-spatial behavior}

Studying the ($k$,$q$) fundamental diagrams 
(Figs.~\ref{fig:VDRTCAKQFundamentalDiagramsForDifferentP} and 
\ref{fig:VDRTCAKVFundamentalDiagramsForDifferentP}) in the previous section, we 
saw the emergence of four distinct density regions (\emph{I})--(\emph{IV}) as $p 
\rightarrow 1.0$ (formed by the $\bigtriangleup$-symbols). In this section, we 
discuss the tempo-spatial properties that are intrinsic to these regions, 
relating them to the previously discussed histograms.\\

As the global density $k$ is increased, the system undergoes three consecutive 
phase transitions (between these four traffic regimes). The densities at which 
these transitions occur, are not just inflection points (such as the finite-size 
effect in the TCA model of Takayasu and Takayasu \cite{BARLOVIC:98}), but they 
signal a new behavior in the system. The next four paragraphs give a detailed 
account of the effects that appear in each density region (i.e., traffic 
regime), as well as the transitions that occur between them.

    \subsubsection{Region (\emph{I}) -- free-flowing traffic [FFT]}
    \label{subsubsec:RegionI}

\begin{figure}[!ht]
  \centering
  \includegraphics[width=8.6cm]{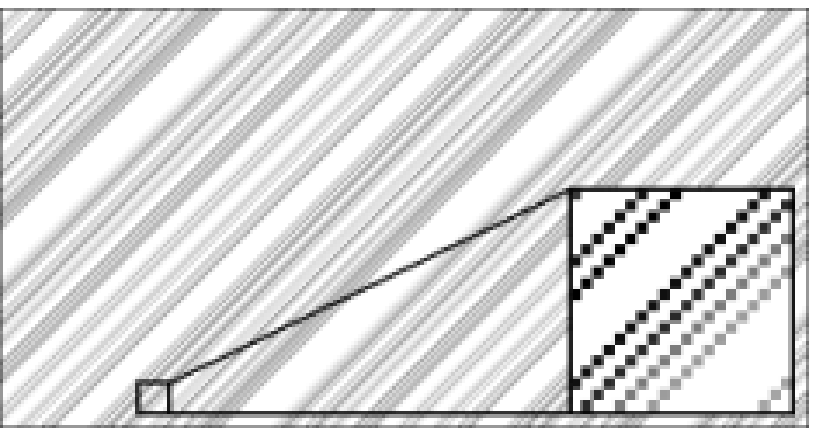}\\
  \vspace*{0.2cm}
  \includegraphics[width=8.6cm]{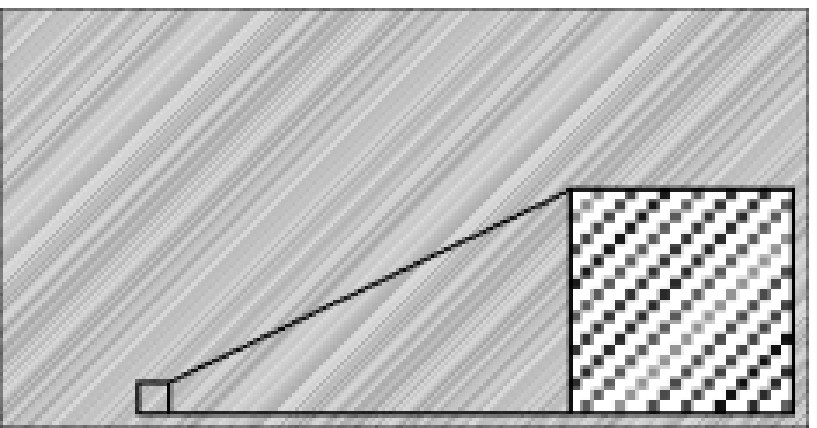}
  \caption{
    Time-space diagrams of the VDR-TCA model with $p_{0} = 0.0$ and $p = 1.0$. 
    The shown lattices each contain 300 cells (corresponding to 2.25~km), the 
    visible time horizon is always 580 seconds. The upper diagram and lower 
    diagrams have densities $k = 0.2$ and $k = \frac{1}{3}$ respectively, 
    corresponding to observations in density region (\emph{I}), labeled 
    \emph{free-flowing traffic} (FFT).
  }
  \label{fig:TXDiagramsVDRTCADensityRegionI}
\end{figure}

For moderately low densities, typical time-space diagrams look as the ones in 
Fig.~\ref{fig:TXDiagramsVDRTCADensityRegionI}. According to the histogram in 
Fig.~\ref{fig:VDRTASpeedHistogramForHighP}, we can see that the speed of all the 
vehicles is the same (namely 1~cell/s) in this density region. Although the 
standard deviation of the space mean speed is zero, this is not the case for the 
average space gap, explaining its rather `nervous' behavior in 
Fig.~\ref{fig:VDRTASpaceGapHistogramForHighP}.

As the global density increases, the transition point between regions (\emph{I}) 
and (\emph{II}) is reached. At this point, each vehicle $i$ has a speed $v_{i} = 
1$~cell/s, and a space gap $g_{s_{i}} = 2$~cells. Because $v_{i} < g_{s_{i}}$, 
case (4a) from section \ref{subsec:QualitativeEffectsRuleSet} applies. 
Using equations (\ref{eq:SpaceHeadway}) and (\ref{eq:SpaceHeadwayAndDensity}), 
we can calculate the corresponding density:

\begin{equation}
  k_{(I) \rightarrow (II)} = \overline h_{s}^{-1} = (\overline L + \overline g_{s})^{-1} = \frac{1}{3}.
\end{equation}

Although the maximum speed any vehicle can (and \emph{will}) travel at is 
limited to 1~cell/s, we still call this state \emph{`free-flowing traffic'} 
(FFT) because no congestion waves are present in the system and none of the 
vehicles has to stop.

    \subsubsection{Region (\emph{II}) -- dilutely congested traffic [DCT]}
    \label{subsubsec:RegionII}

\begin{figure}[!ht]
  \centering
  \includegraphics[width=8.6cm]{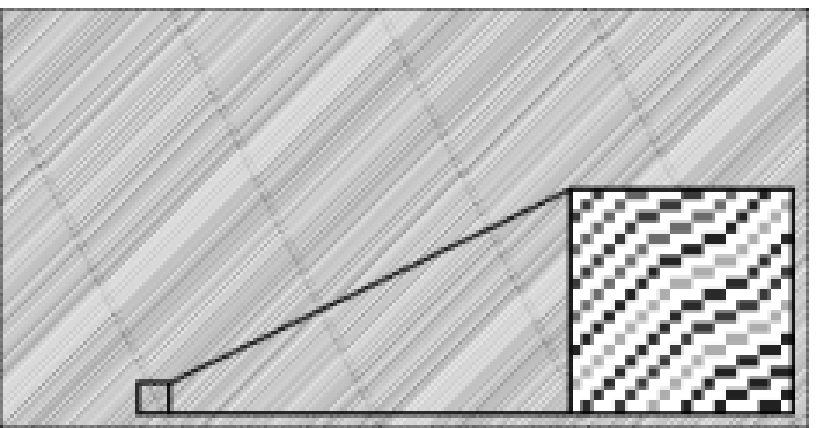}\\
  \vspace*{0.2cm}
  \includegraphics[width=8.6cm]{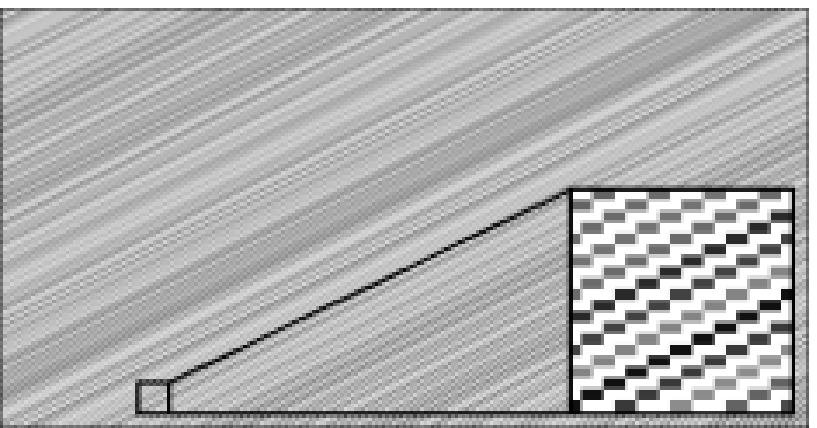}
  \caption{
    Time-space diagrams of the VDR-TCA model with $p_{0} = 0.0$ and $p = 1.0$. 
    The shown lattices each contain 300 cells (corresponding to 2.25~km), the 
    visible time horizon is always 580 seconds. The upper diagram and lower 
    diagrams have densities $k = \frac{1}{3} + \frac{1}{K}$ and $k = 0.4$ 
    respectively, corresponding to observations in density region (\emph{II}), 
    labeled \emph{dilutely congested traffic} (DCT).
  }
  \label{fig:TXDiagramsVDRTCADensityRegionII}
\end{figure}

If we increase the density to $k = \frac{1}{3} + \frac{1}{K}$ (i.e., adding one 
vehicle at the transition point), a new traffic regime is entered. In this 
regime, each extra vehicle leads to a backward propagating mini-jam of 3 
stop-and-go cycles (see Fig.~\ref{fig:TXDiagramsVDRTCADensityRegionII}), 
bringing us to the description of \emph{`dilutely congested traffic'} (DCT).\\

In order to calculate the second transition point, we again observe the 
histogram of Fig.~\ref{fig:VDRTASpaceGapHistogramForHighP}. A vehicle's space 
gap now alternates between 1 and 2~cells in density region (\emph{II}). And 
because its speed alternates between 0 and 1~cell/s respectively, the vehicles' 
motions are controlled by cases (2) and (4a) from section 
\ref{subsec:QualitativeEffectsRuleSet}. This means that stopped vehicles 
accelerate again in the next time step, after which they have to stop again, and 
so indefinitely repeating this cycle of stop-and-go behavior.

Consider now a pair of adjacent driving vehicles $i$ and $j$; it then follows 
from equation (\ref{eq:SpaceHeadway}) that $h_{s_{i}} = 1 + 1 = 2$~cells and 
$h_{s_{j}} = 1 + 2 = 3$~cells. So each pair of vehicles `occupies' 5~cells in 
the lattice, or 2.5~cells on average per vehicle. This leads to the second 
transition point being located at:

\begin{equation}
  k_{(II) \rightarrow (III)} = \overline h_{s}^{-1} = \frac{1}{2.5} = 0.4.
\end{equation}

As the density is increased towards $k_{(II) \rightarrow (III)}$, the space mean 
speed decreases non-linearly. At the transition point itself, $\overline v_{s} = 
0.5$~cells/s and the system is now completely dominated by backward propagating 
dilute jams.

    \subsubsection{Region (\emph{III}) -- densely advancing traffic [DAT]}
    \label{subsubsec:RegionIII}

\begin{figure}[!ht]
  \centering
  \includegraphics[width=8.6cm]{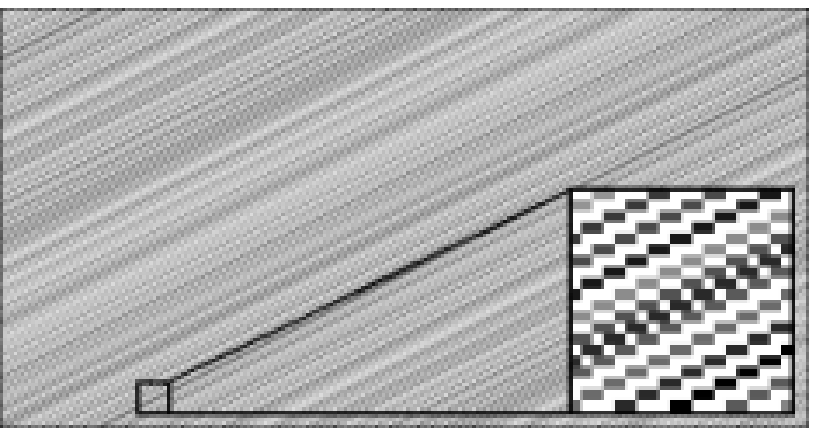}\\
  \vspace*{0.2cm}
  \includegraphics[width=8.6cm]{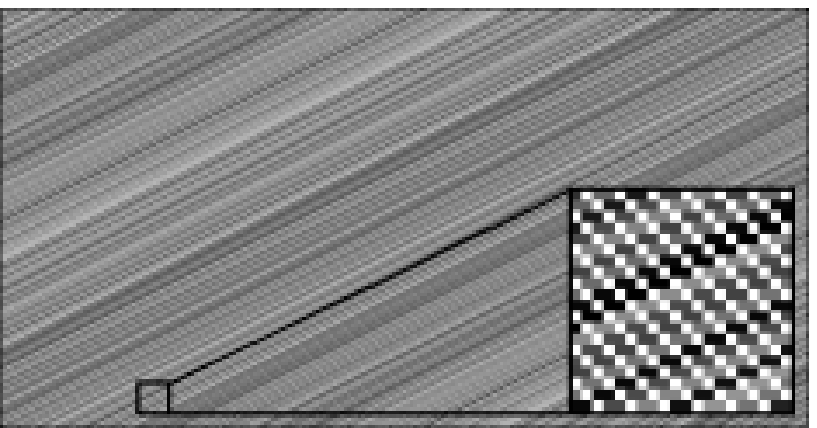}
  \caption{
    Time-space diagrams of the VDR-TCA model with $p_{0} = 0.0$ and $p = 1.0$. 
    The shown lattices each contain 300 cells (corresponding to 2.25~km), the 
    visible time horizon is always 580 seconds. The upper diagram and lower 
    diagrams have densities $k = 0.4 + \frac{1}{K}$ and $k = \frac{2}{3}$ 
    respectively, corresponding to observations in density region (\emph{III}), 
    labeled \emph{densely advancing traffic} (DAT).
}
  \label{fig:TXDiagramsVDRTCADensityRegionIII}
\end{figure}

Adding one more vehicle at the transition point between regions (\emph{II}) and 
(\emph{III}), leads to surprising behavior: a \emph{forward moving jam} emerges, 
traveling at a speed of 0.5~cells/s (see 
Fig.~\ref{fig:TXDiagramsVDRTCADensityRegionIII}). Another artefact is that the 
cycle of alternating space gaps of 1 and 2~cells, is broken with the 
introduction of a zero space gap at the location of this new `jam'. When the 
density reaches the third transition point, the system is completely filled with 
these forward moving `jams' of dense traffic, leading to the description of 
\emph{`densely advancing traffic'} (DAT). Because the available space is more 
optimally used by the vehicles, an increase of the density thus has a 
(temporary) \emph{beneficial} effect on the global flow measured in the system. 
This kind of behavior of forward moving density structures can also be observed 
in some models of anticipatory driving \cite{LARRAGA:04}.

At the transition point itself, all vehicles exhibit the same behavior, 
comparable to the behavior at the second transition point. Traffic is more 
dense, as can be seen in the distribution of the space gaps in 
Fig.~\ref{fig:VDRTASpaceGapHistogramForHighP}: all vehicles have alternatingly 
$g_{s} = 0$ and $g_{s} = 1$~cell (with corresponding speeds of 0 and 1~cell/s 
respectively). One pair of adjacent driving vehicles $i$ and $j$ thus `occupies' 
$h_{s_{i}} + h_{s_{j}} = (1 + 0) + (1 + 1) = 3$~cells, or 1.5~cells on average 
per vehicle. The third transition point thus corresponds to:

\begin{equation}
  k_{(III) \rightarrow (IV)} = \overline h_{s}^{-1} = \frac{1}{1.5} = \frac{2}{3}.
\end{equation}

    \subsubsection{Region (\emph{IV}) -- heavily congested traffic [HCT]}
    \label{subsubsec:RegionIV}

\begin{figure}[!ht]
  \centering
  \includegraphics[width=8.6cm]{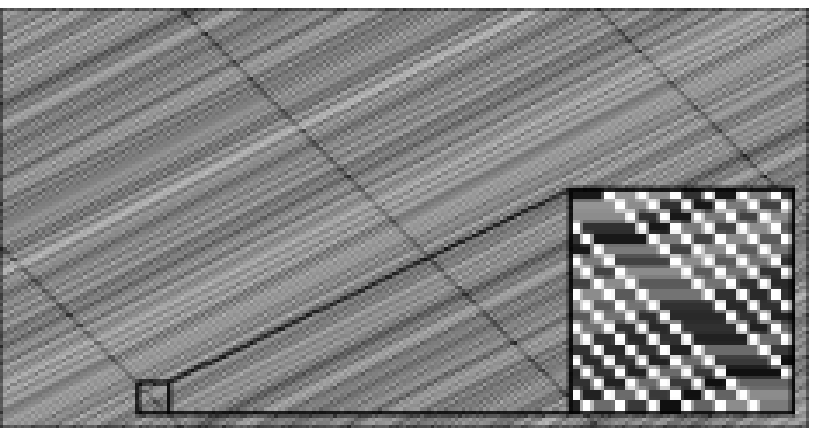}\\
  \vspace*{0.2cm}
  \includegraphics[width=8.6cm]{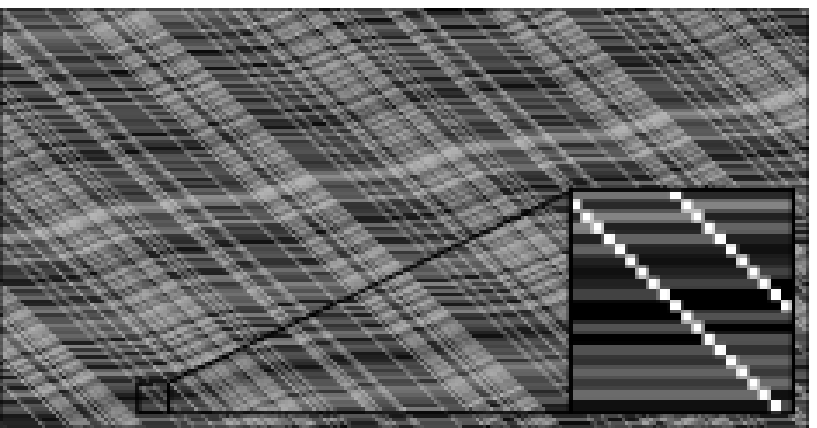}
  \caption{
    Time-space diagrams of the VDR-TCA model with $p_{0} = 0.0$ and $p = 1.0$. 
    The shown lattices each contain 300 cells (corresponding to 2.25~km), the 
    visible time horizon is always 580 seconds. The upper diagram and lower 
    diagrams have densities $k = \frac{2}{3} + \frac{1}{K}$ and $k = 0.85$ 
    respectively, corresponding to observations in density region (\emph{IV}), 
    labeled \emph{heavily congested traffic} (HCT).
  }
  \label{fig:TXDiagramsVDRTCADensityRegionIV}
\end{figure}

Finally, as the system's global density is pushed towards the jam density, each 
extra vehicle introduces at any point in time a backward propagating jam, 
consisting of a block of 5 consecutively stopped vehicles (see 
Fig.~\ref{fig:TXDiagramsVDRTCADensityRegionIV}). The pattern of stable 
stop-and-go traffic gets destroyed, as vehicles remain stopped inside jams for 
longer time periods (cfr. density region (\emph{IV}) in the distribution of the 
time gaps in Fig.~\ref{fig:VDRTATimeGapHistogramForHighP}), leading to the 
description of this traffic regime as \emph{`heavily congested traffic'} (HCT).

  \subsection{Varying the maximum speed}

Up till now, the paper focused on VDR-TCA model's behavior when $p_{0} = 0.0$ 
and $p = 1.0$, for $v_{\mbox{max}} = 5$~cells/s. We can now ask the question: 
\emph{"What would be the effect if we consider a lower maximum speed~?"}\\

\begin{figure}[!ht]
  \centering
  \psfrag{Global density k}[][]{\footnotesize{Global density $k$}}
  \psfrag{Flow q [vehicles/hour]}[][]{\footnotesize{Flow $q$ [vehicles/h]}}
  \psfrag{vMax = 1}[][]{\footnotesize{$v_{\mbox{max}} = 1$}}
  \psfrag{vMax = 2}[][]{\footnotesize{$v_{\mbox{max}} = 2$}}
  \psfrag{vMax = 5}[][]{\footnotesize{$v_{\mbox{max}} = 5$}}
  \includegraphics[width=8.6cm]{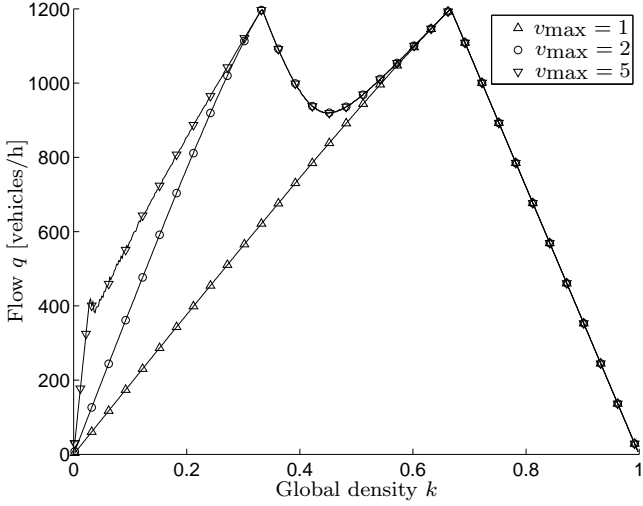}
  \caption{
    The ($k$,$q$) fundamental diagrams of the VDR-TCA with fixed slow-to-start 
    and slowdown probabilities of $p_{0} = 0.0$ and a $p = 0.9$ respectively. 
    The maximum speed was set at 1, 2, and 5~cells/s.
  }
  \label{fig:VDRTCAKQFundamentalDiagramsForDifferentVMax}
\end{figure}

\begin{figure}[!ht]
  \centering
  \psfrag{Global density k}[][]{\footnotesize{Global density $k$}}
  \psfrag{Speed v [cells/s]}[][]{\footnotesize{Space mean speed $\overline v_{s}$ [cells/s]}}
  \psfrag{vMax = 1}[][]{\footnotesize{$v_{\mbox{max}} = 1$}}
  \psfrag{vMax = 2}[][]{\footnotesize{$v_{\mbox{max}} = 2$}}
  \psfrag{vMax = 5}[][]{\footnotesize{$v_{\mbox{max}} = 5$}}
  \includegraphics[width=8.6cm]{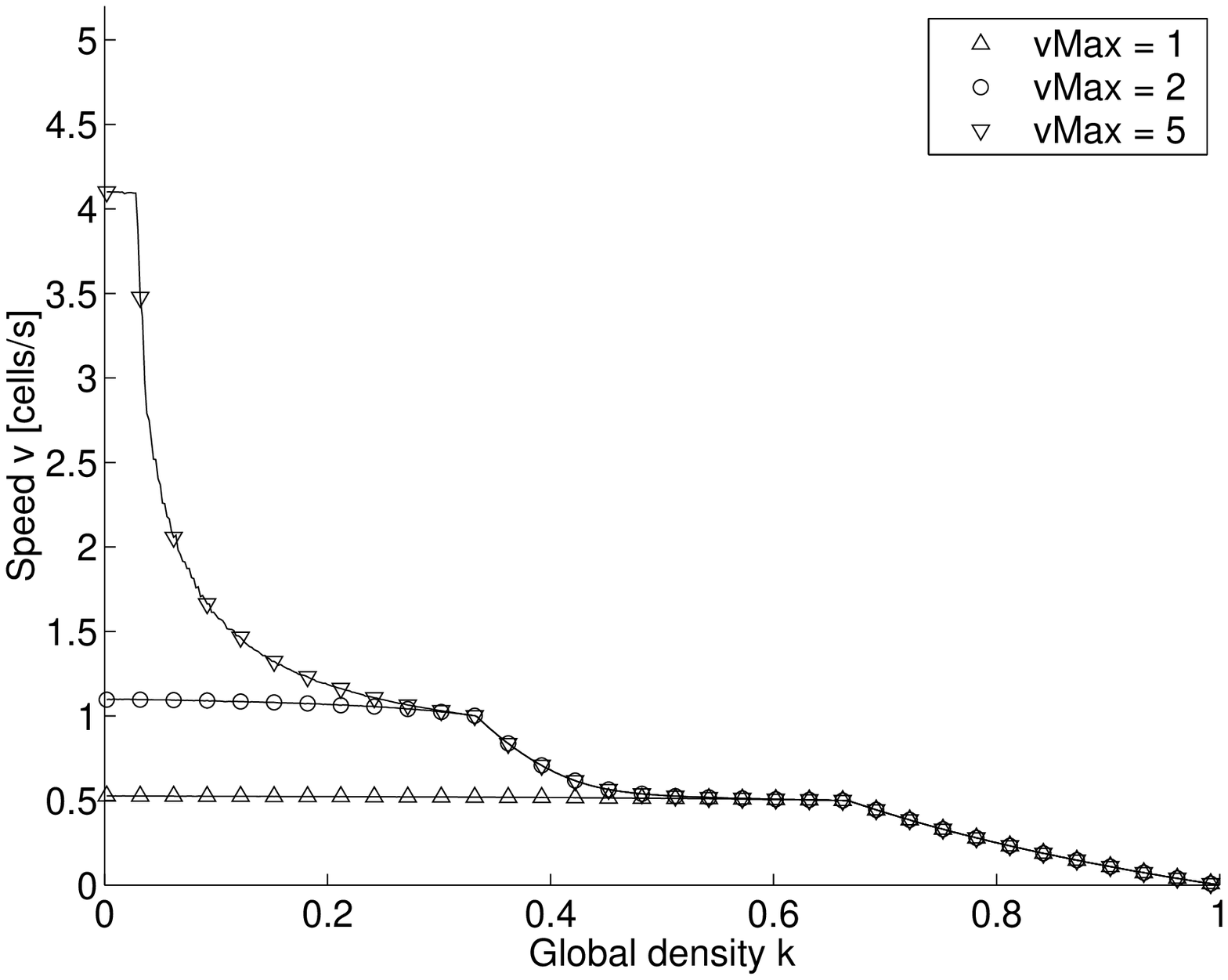}
  \caption{
    The ($k$,$\overline v_{s}$) fundamental diagrams of the VDR-TCA with fixed 
    slow-to-start and slowdown probabilities of $p_{0} = 0.0$ and a $p = 0.9$ 
    respectively. The maximum speed was set at 1, 2, and 5~cells/s.
  }
  \label{fig:VDRTCAKVFundamentalDiagramsForDifferentVMax}
\end{figure}

Considering the ($k$,$q$) fundamental diagrams in 
Fig.~\ref{fig:VDRTCAKQFundamentalDiagramsForDifferentVMax}, we can see that a 
lower maximum speed has the significant effect of seemingly losing the first two 
transitions points $k_{(I) \rightarrow (II)}$ and $k_{(II) \rightarrow (III)}$. 
Put more correctly, when $v_{\mbox{max}} = 1$~cell/s, the first transition point 
$k_{(I) \rightarrow (II)}$ gets drastically shifted towards a higher density of 
$\frac{2}{3}$. A side-effect is that the second and third transitions points 
vanish. This effect is also visible in the ($k$,$\overline v_{s}$) fundamental 
diagrams in Fig.~\ref{fig:VDRTCAKVFundamentalDiagramsForDifferentVMax}. Note 
that we have set $p = 0.9$ in order to make the differences between the 
fundamental diagrams more pronounced.

A decrease of the maximum speed thus has only a significant effect in the 
limiting case where $v_{\mbox{max}} = 1$~cell/s, leading to a less rich 
dynamical behavior in the system. Any other maximum speed $v_{\mbox{max}} > 
1$~cell/s gives rise to different traffic regimes. Note that all the branches in 
the congested regions of the fundamental diagrams coincide once the first 
transition between regions (\emph{I}) and (\emph{II}) occurs at $k_{(I) 
\rightarrow (II)} = \frac{1}{3}$.

  \subsection{Tracking the phase transitions}

In order to track the transitions that occur between the different regimes 
(\emph{I})--(\emph{IV}), we look for a suitable order parameter that expresses a 
qualitative behavior that changes between these regimes. In the following 
paragraphs, we describe the use of two such parameters: the first one is based 
on correlations between neighboring cells, whereas the second one is based on a 
comparison of the difference between the system's global and local densities.

    \subsubsection{Nearest neighbor correlations}

One example of such a suitable measure, is the nearest neighbor order parameter 
$M_{1}$, which gives the time-space averaged density of those vehicles with 
speed $v = 0$~cells/s which had to brake due to the next vehicle ahead 
\cite{EISENBLATTER:98,SCHADSCHNEIDER:02}.

\begin{figure}[!ht]
  \centering
  \psfrag{Global density k}[][]{\footnotesize{Global density $k$}}
  \psfrag{Order parameter M1}[][]{\footnotesize{Order parameter $M_{1}$}}
  \psfrag{p0 = 0.01  p = 0.50}[][]{\footnotesize{$p_{0} = 0.01$  $p = 0.5$}}
  \psfrag{p0 = 0.00  p = 1.00}[][]{\footnotesize{$p_{0} = 0.0$  $p = 1.0$}}
  \includegraphics[width=8.6cm]{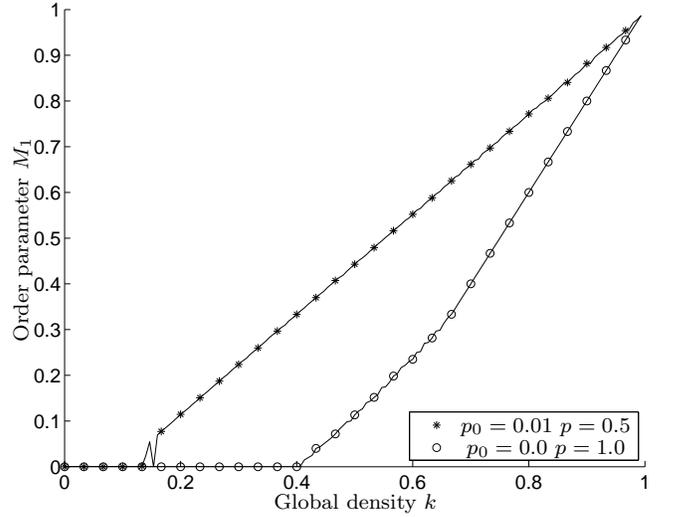}
  \caption{
    The $M_{1}$ nearest neighbor order parameter for the two systems with 
    $(p_{0},p) = (0.5,0.01)$ and $(p_{0},p) = (0.0,1.0)$ respectively. The 
    system size was 300 cells and the total simulation period $T_{\mbox{sim}} = 
    10^{5}$~s.
  }
  \label{fig:NNOrderParameter}
\end{figure}

If we define the occupancy $\eta_{i}$ of cell $i$ as:

\begin{equation}
\left \lbrace
  \begin{array}{lcl}
    \eta_{i} = 1 & \Leftrightarrow & \mbox{cell $i$ is occupied by a vehicle},\\
    \eta_{i} = 0 & \Leftrightarrow & \mbox{cell $i$ is empty},
  \end{array}
\right.
\end{equation}

\noindent
then we can define the order parameter $M_{1}$ as the following function that 
captures the density correlation between pairs of nearest neighbors (at a 
certain global density $k$):

\begin{equation}
\label{eq:OrderParameterM1}
  M_{1} = \left < \frac{1}{K} \sum_{i = 1}^{K} \eta_{i}\eta_{i + 1} \right >_{T_{\mbox{sim}}},
\end{equation}

\noindent
with $\eta_{K + 1} \equiv \eta_{1}$. Note that equation 
(\ref{eq:OrderParameterM1}) is calculated as an average over the total 
simulation period $T_{\mbox{sim}}$. Figure~\ref{fig:NNOrderParameter} shows the 
results from an experiment for calculating $M_{1}$ when $K = 300$~cells, 
$v_{\mbox{max}} = 5$~cells/s and $T_{\mbox{sim}} = 10^{5}$~s (with a transient 
period of $10^{3}$~s).\\

For the normal case where $p_{0} = 0.5$ and $p = 0.01$, we can see that $M_{1}$ 
is clearly able to detect the transition at the critical density: the order 
parameter shows a sudden significant increase. Although the VDR-TCA model 
exhibits metastability, it has only two distinct phases, namely free-flowing and 
jammed. These two phases are detected by the order parameter as it is constantly 
zero in the former, whereas it's a linearly increasing function of the density 
in the latter. As stated, in the free-flowing regime, the nearest neighbor 
correlations are always zero because all the vehicles have large enough space 
gaps (i.e., at least the cell immediately in front of a vehicle is always 
empty).

If we consider the behavior of the order parameter $M_{1}$ in the second case, 
where $p_{0} = 0.0$ and $p = 1.0$, we see that it is \emph{not} able to detect 
the first transition at $k_{(I) \rightarrow (II)} = \frac{1}{3}$. The reason for 
this can be determined by looking at the tempo-spatial evolutions in 
Figs.~\ref{fig:TXDiagramsVDRTCADensityRegionI} and 
\ref{fig:TXDiagramsVDRTCADensityRegionII}. Here we can see that, as described in 
sections \ref{subsubsec:RegionI} and \ref{subsubsec:RegionII}, each vehicle 
\emph{always} has at least one empty cell in front of it, leading to a nearest 
neighbor density correlation of zero.

The order parameter is however succesful in detecting the second transition at 
$k_{(II) \rightarrow (III)} = 0.4$. The third transition at $k_{(III) 
\rightarrow (IV)} = \frac{2}{3}$ is detected, but in a non-intuitive manner: 
only the slope of the linear dependency between $M_{1}$ and $k$ changes when 
going from region (\emph{III}) to region (\emph{IV}).\\

We conclude this paragraph by stating that the $M_{1}$ order parameter is able 
to capture \emph{some} transitions between different traffic regimes, but not 
all of them. Furthermore, it seems that the ability of the parameter to 
distinguish between two regimes is qualitively inferior: this can be witnessed 
by the weak behavioral change at the transition point $k_{(III) \rightarrow 
(IV)}$.

    \subsubsection{A measure of inhomogeneity}

A more suitable order parameter that can accurately track the transitions 
between the different regimes, is the order parameter $M_{2}$ defined by Jost 
and Nagel \cite{JOST:03}. They partition the road into $L$ equal segments of 
integral length $K/L$. After calculating the \emph{local} densities $k_{l_{i}}$ 
for all these segments $i$, they consider the variance of the local densities as 
their order parameter :

\begin{equation}
\label{eq:OrderParameterM2}
  M_{2} = \left < \frac{1}{L} \sum_{i = 1}^{L} (k_{l_{i}} - \mbox{E}[k_{l}])^{2} \right >_{T_{\mbox{sim}}}.
\end{equation}

In equation (\ref{eq:OrderParameterM2}), $\mbox{E}[k_{l}]$ denotes the expected 
value of the local density, which is equal to the global density $k$ because we 
consider a closed system. The local densities are calculated as:

\begin{equation}
  k_{l_{i}} = \frac{1}{K / L} \sum_{j = i_{1}}^{j = i_{K / L}} \eta_{j},
\end{equation}

\noindent
with $i_{1}$ and $i_{K / L}$ denoting respectively the beginning and ending of 
segment $i$. The order parameter $M_{2}$ thus effectively tracks the differences 
in densities of individual segments, leading to a measure for the inhomogeneity 
in the system.\\

Note that $M_{2}$ is calculated as an average over $T_{\mbox{sim}}$ time steps, 
just as was done for the previous order parameter $M_{1}$. This means that we 
calculated the summation in equation (\ref{eq:OrderParameterM2}) at each time 
step, and averaged all these results over a simulation period of 
$T_{\mbox{sim}}$. Care should be taken not to prematurely average the local 
densities over more than one time step, as this would significantly deteriorate 
the tracking quality of the order parameter: the longer the aggregation period 
for calculating the local densities, the more each segment `sees' the densities 
recorded in the other segments as congestion waves are traveling around the 
closed system.\\

\begin{figure}[!ht]
  \centering
  \psfrag{Global density k}[][]{\footnotesize{Global density $k$}}
  \psfrag{Inhomogeneity measure M2}[][]{\footnotesize{Inhomogeneity measure $M_{2}$}}
  \psfrag{p = 0.10}[][]{\footnotesize{$p = 0.10$}}
  \psfrag{p = 0.25}[][]{\footnotesize{$p = 0.25$}}
  \psfrag{p = 0.50}[][]{\footnotesize{$p = 0.50$}}
  \psfrag{p = 0.75}[][]{\footnotesize{$p = 0.75$}}
  \psfrag{p = 0.90}[][]{\footnotesize{$p = 0.90$}}
  \psfrag{p = 1.00}[][]{\footnotesize{$p = 1.00$}}
  \includegraphics[width=8.6cm]{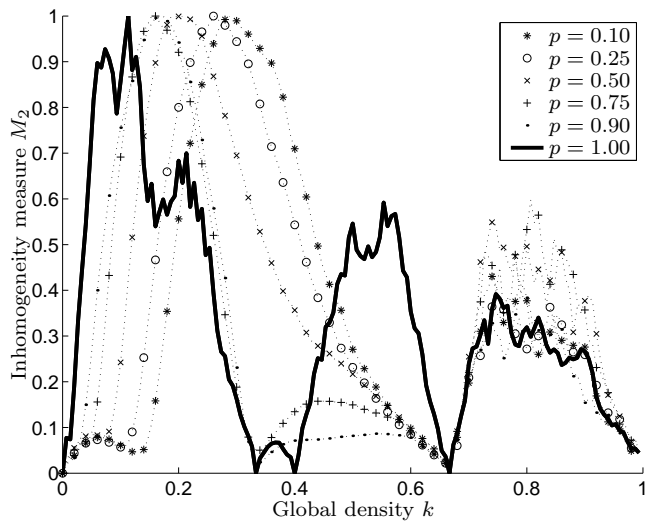}
  \caption{
    The $M_{2}$ order parameter which measures the variance of the local 
    densities. In each diagram, the slow-to-start probability $p_{0}$ was fixed 
    at 0.0, whereas the slowdown probability $p$ was varied between 0.1 and 1.0 
    (the system size was 300 cells and the total simulation period 
    $T_{\mbox{sim}} = 10^{5}$~s). The thick line denotes the limiting case where 
    $p_{0} = 0.0$ and $p = 1.0$.
  }
  \label{fig:OrderParameterM2}
\end{figure}

In Fig.~\ref{fig:OrderParameterM2} we can see the results for different slowdown 
probabilities $p$ (the slow-to-start probability $p_{0}$ was fixed at 0.0). The 
system size $K$ was set to 300 cells, $v_{\mbox{max}} = 5$~cells/s and 
$T_{\mbox{sim}} = 10^{5}$~s (with a transient period of $10^{3}$~s).

From the graph of $M_{2}$ with $p_{0} = 0.0$ and $p = 1.0$, it is interesting to 
notice that this order parameter performs exquisitely well at tracking the 
transitions between the different traffic regimes. Each time a transition point 
is encountered, $M_{2}$ drops to zero. This specific behavior of the order 
parameter can be understood by looking at the tempo-spatial conditions that hold 
at these transitional densities: each time, the system settles in a complete 
homogeneous equilibrium (as can be seen from the lower time-space diagrams in 
Figs.~\ref{fig:TXDiagramsVDRTCADensityRegionI}, 
\ref{fig:TXDiagramsVDRTCADensityRegionII} and 
\ref{fig:TXDiagramsVDRTCADensityRegionIII}). The distribution of the vehicles' 
space gaps (Fig.~\ref{fig:VDRTASpaceGapHistogramForHighP}) also has a variance 
of zero at the first transition point, but not at the others. We thus conclude 
that this inhomogeneity measure is able to accurately track the four previously 
discussed phase transitions.

\section{Comparison with existing literature}
\label{section:LiteratureStudy}

A final word should be said about the behavior of the four phases discussed in 
this paper. Although the study is based on a well-known traffic flow model 
(i.e., the VDR-TCA model), it seems that this behavior is \emph{drastically 
different} from that observed in real-life traffic. The research elucidated in 
this paper, is therefore relevant as it can be compared to phase transitions in 
other types of granular media, described by cellular automata.\\

Relating our findings to similar considerations in literature, we note the 
following aspects: Grabolus \cite{GRABOLUS:01} gives a thorough discussion of 
several variants of the STCA, including so-called `fast-to-start' TCA models 
that give rise to forward propagating density waves; the study refers to the 
two-fluid theory \cite{WILLIAMS:97} as a possible explanation of this 
phenomenon. Gray and Griffeath \cite{GRAY:01} discuss the origin of 
non-concavity in the ($k$,$q$) fundamental diagram, where they propose an 
explanation in which the rear end of a jam is unstable, whereas its front is 
stable and growing. Leveque \cite{LEVEQUE:01} incorporates a non-concave 
fundamental diagram itself in a macroscopic model, thereby resembling night 
driving. Nishinari et al \cite{NISHINARI:03} consider the tempo-spatial 
organization of ants using an ant trail model based on pheromones. In their 
model, the average speed of the ants varies non-monotonically with their 
density, leading to an inflection point in the ($k$,$q$) fundamental diagram. 
Zhang \cite{ZHANG:03} questions the property of anisotropy in \emph{multi-lane} 
traffic, leading to strikingly similar non-concave ($k$,$q$) fundamental 
diagrams.\\

Finally, Awazu \cite{AWAZU:99} investigated the flow of various complex 
particles using a simple meta-model based on Wolfram's CA-rule 184, but extended 
with rules that govern the change in speed of individual particles. As a result, 
he classified three types of fundamental diagrams, namely two phases (2P), three 
phases (3P), and four phases (4P) type systems. Of these types, the 4P-type 
complex granular particle systems closely resemble the regimes discovered in 
this paper. Awazu discusses the types of regimes and the transitions between 
them, using the terminology of `dilute slugs' for the slow moving jams in the 
DCT-phase (see section \ref{subsubsec:RegionII}), which he calls the `dilute 
jam-flow state'. Analogously, there are `advancing slugs' in the `advancing 
jam-flow state' (related to our DAT-phase in section \ref{subsubsec:RegionIII}) 
and the fourth phase is called the `hard jam-flow state' (related to our 
HCT-phase in section \ref{subsubsec:RegionIV}).

As stated before, the behavior here is different from that of real-life traffic 
flows. Awazu expects that these transitions appear if there are more complex 
interactions between the particles. In the 4P-type systems, attractive forces 
between close neighbouring particles and an effective resistance on moving 
neighbouring particles, lead to the realization of the previously discussed flow 
phases.

\section{Summary}
\label{section:Summary}

In this paper, we first showed the behavioral characteristics resulting from the 
velocity dependent randomization traffic cellular automaton (VDR-TCA) model, 
operating under normal conditions (i.e., $p_{0} \gg p$). Then we investigated 
the more complex system dynamics that arise from this traffic flow model in the 
exceptional cases when $p_{0} \ll p$. This behavior was quantitavely compared 
against the VDR-TCA's normal operation, using classical fundamental diagrams and 
histograms that show the distribution of the speeds, space, and time gaps.\\

Our main investigations were primarily directed at the limiting case where 
$p_{0} = 0.0$ and $p = 1.0$. We discovered the emergence of four different 
traffic regimes. These regimes were individually studied using diagrams that 
show the evolution of their tempo-spatial behavioral characteristics. This 
resulted in the following classification: free-flowing traffic (FFT), dilutely 
congested traffic (DCT), densely advancing traffic (DAT), and heavily congested 
traffic (HCT). Our main conclusions here are:

\begin{itemize}
  \item all four phases share the common property that moving vehicles can 
    never increase their speed once the system has settled into an equilibrium, 
  \item the DAT regime experiences forward propagating density waves, 
    corresponding to a non-concave region in the system's flow-density relation.
\end{itemize}

In order to track the phase transitions between these traffic regimes, we looked 
at two order parameters: the first one $M_{1}$ was based on correlations between 
neighboring cells, whereas the second one $M_{2}$ was based on a comparison of 
the difference between the system's global and local densities. We concluded 
that, as opposed to $M_{1}$, this latter order parameter $M_{2}$ performs 
exemplary at tracking the transitions between the different traffic regimes.\\

Comparing our results with those in existing literature, we conclude that the 
work of Awazu \cite{AWAZU:99}, dealing with a cellular automaton model of 
various complex particles, gives the closest resemblance to our four phases.

\section*{Acknowledgements}

S. Maerivoet would like to thank dr. Andreas Schadschneider for his insightful 
comments and feedback when writing this paper.\\

Our research is supported by:\\
\textbf{Research Council KUL}: GOA-Mefisto 666, several PhD/postdoc \& fellow 
grants,\\
\textbf{FWO}: PhD/postdoc grants, projects, G.0240.99 (multilinear algebra), 
G.0407.02 (support vector machines), G.0197.02 (power islands), G.0141.03 
(Identification and cryptography), G.0491.03 (control for intensive care 
glycemia), G.0120.03 (QIT), G.0800.01 (collective intelligence), G.0452.04 (new 
quantum algorithms), G.0499.04 (kernel based methods), research communities 
(ICCoS, ANMMM),
\textbf{AWI}: Bil. Int. Collaboration Hungary/Poland,\\
\textbf{IWT}: PhD Grants,\\
\textbf{Belgian Federal Science Policy Office}: IUAP P5/22 (`Dynamical Systems 
and Control: Computation, Identification and Modelling'), PODO-II (CP/40: TMS 
and sustainability),\\
\textbf{EU}: FP5-CAGE, FP5-Quprodis, ERNSI, FP6-BioPattern, Eureka 2419-FliTE,\\
\textbf{Contract Research/agreements}: Data4s, Electrabel, Elia, LMS, IPCOS, 
VIB.

\bibliography{paper}

\begin{thebibliography}{27}
\expandafter\ifx\csname natexlab\endcsname\relax\def\natexlab#1{#1}\fi
\expandafter\ifx\csname bibnamefont\endcsname\relax
  \def\bibnamefont#1{#1}\fi
\expandafter\ifx\csname bibfnamefont\endcsname\relax
  \def\bibfnamefont#1{#1}\fi
\expandafter\ifx\csname citenamefont\endcsname\relax
  \def\citenamefont#1{#1}\fi
\expandafter\ifx\csname url\endcsname\relax
  \def\url#1{\texttt{#1}}\fi
\expandafter\ifx\csname urlprefix\endcsname\relax\def\urlprefix{URL }\fi
\providecommand{\bibinfo}[2]{#2}
\providecommand{\eprint}[2][]{\url{#2}}

\bibitem[{\citenamefont{Maerivoet and Moor}(2003)}]{MAERIVOET:03d}
\bibinfo{author}{\bibfnamefont{S.}~\bibnamefont{Maerivoet}} \bibnamefont{and}
  \bibinfo{author}{\bibfnamefont{B.~D.} \bibnamefont{Moor}}, in
  \emph{\bibinfo{booktitle}{Proceedings of the 10th World Congress and
  Exhibition on Intelligent Transport Systems and Services (CD-ROM)}}
  (\bibinfo{organization}{ERTICO, ITS Europe}, \bibinfo{address}{Madrid,
  Spain}, \bibinfo{year}{2003}).

\bibitem[{\citenamefont{Nagel and Schreckenberg}(1992)}]{NAGEL:92}
\bibinfo{author}{\bibfnamefont{K.}~\bibnamefont{Nagel}} \bibnamefont{and}
  \bibinfo{author}{\bibfnamefont{M.}~\bibnamefont{Schreckenberg}},
  \bibinfo{journal}{Journal de Physique I France} \textbf{\bibinfo{volume}{2}},
  \bibinfo{pages}{2221} (\bibinfo{year}{1992}).

\bibitem[{\citenamefont{Nagel}(1994)}]{NAGEL:94b}
\bibinfo{author}{\bibfnamefont{K.}~\bibnamefont{Nagel}},
  \bibinfo{journal}{International Journal of Modern Physics C}
  \textbf{\bibinfo{volume}{5}}, \bibinfo{pages}{567} (\bibinfo{year}{1994}).

\bibitem[{\citenamefont{Takayasu and Takayasu}(1993)}]{TAKAYASU:93}
\bibinfo{author}{\bibfnamefont{M.}~\bibnamefont{Takayasu}} \bibnamefont{and}
  \bibinfo{author}{\bibfnamefont{H.}~\bibnamefont{Takayasu}},
  \bibinfo{journal}{Fractals 1} pp. \bibinfo{pages}{860--866}
  (\bibinfo{year}{1993}).

\bibitem[{\citenamefont{Barlovi{\'c} et~al.}(1998)\citenamefont{Barlovi{\'c},
  Santen, Schadschneider, and Schreckenberg}}]{BARLOVIC:98}
\bibinfo{author}{\bibfnamefont{R.}~\bibnamefont{Barlovi{\'c}}},
  \bibinfo{author}{\bibfnamefont{L.}~\bibnamefont{Santen}},
  \bibinfo{author}{\bibfnamefont{A.}~\bibnamefont{Schadschneider}},
  \bibnamefont{and}
  \bibinfo{author}{\bibfnamefont{M.}~\bibnamefont{Schreckenberg}},
  \bibinfo{journal}{European Physics Journal} \textbf{\bibinfo{volume}{B5}}
  (\bibinfo{year}{1998}).

\bibitem[{\citenamefont{Awazu}(1999)}]{AWAZU:99}
\bibinfo{author}{\bibfnamefont{A.}~\bibnamefont{Awazu}},
  \bibinfo{journal}{Physics Letters A} \textbf{\bibinfo{volume}{261}},
  \bibinfo{pages}{309} (\bibinfo{year}{1999}).

\bibitem[{\citenamefont{Grabolus}(2001)}]{GRABOLUS:01}
\bibinfo{author}{\bibfnamefont{S.}~\bibnamefont{Grabolus}}, Master's thesis,
  \bibinfo{school}{Institut f{\"u}r Theoretische Physik, Universit{\"a}t zu
  K{\"o}ln} (\bibinfo{year}{2001}).

\bibitem[{\citenamefont{Gray and Griffeath}(2001)}]{GRAY:01}
\bibinfo{author}{\bibfnamefont{L.}~\bibnamefont{Gray}} \bibnamefont{and}
  \bibinfo{author}{\bibfnamefont{D.}~\bibnamefont{Griffeath}},
  \bibinfo{journal}{Journal of Statistical Physics}
  \textbf{\bibinfo{volume}{105}}, \bibinfo{pages}{413} (\bibinfo{year}{2001}).

\bibitem[{\citenamefont{Leveque}(2001)}]{LEVEQUE:01}
\bibinfo{author}{\bibfnamefont{R.~J.} \bibnamefont{Leveque}}, in
  \emph{\bibinfo{booktitle}{Minisymposium on traffic flow,}}
  (\bibinfo{organization}{SIAM Annual Meeting}, \bibinfo{year}{2001}).

\bibitem[{\citenamefont{Nishinari et~al.}(2003)\citenamefont{Nishinari,
  Chowdhury, and Schadschneider}}]{NISHINARI:03}
\bibinfo{author}{\bibfnamefont{K.}~\bibnamefont{Nishinari}},
  \bibinfo{author}{\bibfnamefont{D.}~\bibnamefont{Chowdhury}},
  \bibnamefont{and}
  \bibinfo{author}{\bibfnamefont{A.}~\bibnamefont{Schadschneider}},
  \bibinfo{journal}{Physical Review E} \textbf{\bibinfo{volume}{67}},
  \bibinfo{pages}{036120} (\bibinfo{year}{2003}).

\bibitem[{\citenamefont{Zhang}(2003)}]{ZHANG:03}
\bibinfo{author}{\bibfnamefont{H.}~\bibnamefont{Zhang}},
  \bibinfo{journal}{Transportation Research B} \textbf{\bibinfo{volume}{37B}},
  \bibinfo{pages}{561} (\bibinfo{year}{2003}).

\bibitem[{\citenamefont{Nagel}(1995)}]{NAGEL:95c}
\bibinfo{author}{\bibfnamefont{K.}~\bibnamefont{Nagel}}, Ph.D. thesis,
  \bibinfo{school}{Universit{\"a}t zu K{\"o}ln} (\bibinfo{year}{1995}).

\bibitem[{\citenamefont{Chowdhury et~al.}(1998)\citenamefont{Chowdhury,
  Pasupathy, and Sinha}}]{CHOWDHURY:98}
\bibinfo{author}{\bibfnamefont{D.}~\bibnamefont{Chowdhury}},
  \bibinfo{author}{\bibfnamefont{A.}~\bibnamefont{Pasupathy}},
  \bibnamefont{and} \bibinfo{author}{\bibfnamefont{S.}~\bibnamefont{Sinha}},
  \bibinfo{journal}{European Physics Journal B - Condensed Matter}
  \textbf{\bibinfo{volume}{5}}, \bibinfo{pages}{781} (\bibinfo{year}{1998}).

\bibitem[{\citenamefont{Schadschneider}(2000)}]{SCHADSCHNEIDER:00}
\bibinfo{author}{\bibfnamefont{A.}~\bibnamefont{Schadschneider}},
  \bibinfo{journal}{Physica A} p. \bibinfo{pages}{101} (\bibinfo{year}{2000}).

\bibitem[{\citenamefont{Helbing}(2001)}]{HELBING:01}
\bibinfo{author}{\bibfnamefont{D.}~\bibnamefont{Helbing}},
  \bibinfo{journal}{Review of Modern Physics} \textbf{\bibinfo{volume}{73}},
  \bibinfo{pages}{1067} (\bibinfo{year}{2001}).

\bibitem[{\citenamefont{Maerivoet}(2004)}]{MAERIVOET:04d}
\bibinfo{author}{\bibfnamefont{S.}~\bibnamefont{Maerivoet}},
  \emph{\bibinfo{title}{\emph{Traffic Cellular Automata}}},
  \bibinfo{howpublished}{Java software tested with JDK 1.3.1}
  (\bibinfo{year}{2004}), \bibinfo{note}{\\(URL~:
  \texttt{http://smtca.dyns.cx})}.

\bibitem[{\citenamefont{Chowdhury et~al.}(2000)\citenamefont{Chowdhury, Santen,
  and Schadschneider}}]{CHOWDHURY:00}
\bibinfo{author}{\bibfnamefont{D.}~\bibnamefont{Chowdhury}},
  \bibinfo{author}{\bibfnamefont{L.}~\bibnamefont{Santen}}, \bibnamefont{and}
  \bibinfo{author}{\bibfnamefont{A.}~\bibnamefont{Schadschneider}},
  \bibinfo{journal}{Physics Reports} \textbf{\bibinfo{volume}{329}},
  \bibinfo{pages}{199} (\bibinfo{year}{2000}).

\bibitem[{\citenamefont{Barlovi{\'c} et~al.}(1999)\citenamefont{Barlovi{\'c},
  Esser, Froese, Knospe, Neubat, Schreckenberg, and Wahle}}]{BARLOVIC:99}
\bibinfo{author}{\bibfnamefont{R.}~\bibnamefont{Barlovi{\'c}}},
  \bibinfo{author}{\bibfnamefont{J.}~\bibnamefont{Esser}},
  \bibinfo{author}{\bibfnamefont{K.}~\bibnamefont{Froese}},
  \bibinfo{author}{\bibfnamefont{W.}~\bibnamefont{Knospe}},
  \bibinfo{author}{\bibfnamefont{L.}~\bibnamefont{Neubat}},
  \bibinfo{author}{\bibfnamefont{M.}~\bibnamefont{Schreckenberg}},
  \bibnamefont{and} \bibinfo{author}{\bibfnamefont{J.}~\bibnamefont{Wahle}},
  \bibinfo{journal}{Traffic and Mobility~: Simulation-Economics-Environment}
  pp. \bibinfo{pages}{117--134} (\bibinfo{year}{1999}),
  \bibinfo{note}{{Institut f{\"u}r Kraftfahrwesen, RWTH Aachen, Duisburg}}.

\bibitem[{\citenamefont{Wolfram}(2002)}]{WOLFRAM:02}
\bibinfo{author}{\bibfnamefont{S.}~\bibnamefont{Wolfram}},
  \emph{\bibinfo{title}{A New Kind of Science}} (\bibinfo{publisher}{Wolfram
  Media, Inc.}, \bibinfo{year}{2002}), \bibinfo{note}{{ISBN 1-579-955008-8}}.

\bibitem[{\citenamefont{Schadschneider}(2002)}]{SCHADSCHNEIDER:02}
\bibinfo{author}{\bibfnamefont{A.}~\bibnamefont{Schadschneider}},
  \bibinfo{journal}{Physica A} \textbf{\bibinfo{volume}{313}},
  \bibinfo{pages}{153} (\bibinfo{year}{2002}).

\bibitem[{\citenamefont{Maerivoet et~al.}(2003)\citenamefont{Maerivoet, Logghe,
  Moor, and Immers}}]{MAERIVOET:03g}
\bibinfo{author}{\bibfnamefont{S.}~\bibnamefont{Maerivoet}},
  \bibinfo{author}{\bibfnamefont{S.}~\bibnamefont{Logghe}},
  \bibinfo{author}{\bibfnamefont{B.~D.} \bibnamefont{Moor}}, \bibnamefont{and}
  \bibinfo{author}{\bibfnamefont{B.}~\bibnamefont{Immers}}, in
  \emph{\bibinfo{booktitle}{Proceedings of the Workshop on Traffic and Granular
  Flow '03}} (\bibinfo{organization}{Delft University of Technology},
  \bibinfo{address}{Delft, The Netherlands}, \bibinfo{year}{2003}).

\bibitem[{\citenamefont{L{\'a}rraga et~al.}(2004)\citenamefont{L{\'a}rraga, del
  R{\'i}o, and Schadschneider}}]{LARRAGA:04}
\bibinfo{author}{\bibfnamefont{M.}~\bibnamefont{L{\'a}rraga}},
  \bibinfo{author}{\bibfnamefont{J.}~\bibnamefont{del R{\'i}o}},
  \bibnamefont{and}
  \bibinfo{author}{\bibfnamefont{A.}~\bibnamefont{Schadschneider}},
  \bibinfo{journal}{Journal of Physics A: Mathematical and General} pp.
  \bibinfo{pages}{3769--3781} (\bibinfo{year}{2004}).

\bibitem[{\citenamefont{Schreckenberg et~al.}(2001)\citenamefont{Schreckenberg,
  R.Barlovi{\'c}, Knospe, and Kl{\"u}pfel}}]{SCHRECKENBERG:01}
\bibinfo{author}{\bibfnamefont{M.}~\bibnamefont{Schreckenberg}},
  \bibinfo{author}{\bibnamefont{R.Barlovi{\'c}}},
  \bibinfo{author}{\bibfnamefont{W.}~\bibnamefont{Knospe}}, \bibnamefont{and}
  \bibinfo{author}{\bibfnamefont{H.}~\bibnamefont{Kl{\"u}pfel}}, in
  \emph{\bibinfo{booktitle}{Computational Statistical Physics}}, edited by
  \bibinfo{editor}{\bibfnamefont{K.~H.} \bibnamefont{Hoffmann}}
  \bibnamefont{and} \bibinfo{editor}{\bibfnamefont{M.}~\bibnamefont{Schreiber}}
  (\bibinfo{publisher}{Springer}, \bibinfo{address}{Berlin},
  \bibinfo{year}{2001}), pp. \bibinfo{pages}{113--126}.

\bibitem[{\citenamefont{Eisenbl{\"a}tter
  et~al.}(1998)\citenamefont{Eisenbl{\"a}tter, Santen, Schadschneider, and
  Schreckenberg}}]{EISENBLATTER:98}
\bibinfo{author}{\bibfnamefont{B.}~\bibnamefont{Eisenbl{\"a}tter}},
  \bibinfo{author}{\bibfnamefont{L.}~\bibnamefont{Santen}},
  \bibinfo{author}{\bibfnamefont{A.}~\bibnamefont{Schadschneider}},
  \bibnamefont{and}
  \bibinfo{author}{\bibfnamefont{M.}~\bibnamefont{Schreckenberg}},
  \bibinfo{journal}{Physical Review E} \textbf{\bibinfo{volume}{57}},
  \bibinfo{pages}{1309} (\bibinfo{year}{1998}).

\bibitem[{\citenamefont{Jost and Nagel}(2003)}]{JOST:03}
\bibinfo{author}{\bibfnamefont{D.}~\bibnamefont{Jost}} \bibnamefont{and}
  \bibinfo{author}{\bibfnamefont{K.}~\bibnamefont{Nagel}}, in
  \emph{\bibinfo{booktitle}{Transportation Research Board Annual Meeting}}
  (\bibinfo{address}{Washington, D.C.}, \bibinfo{year}{2003}),
  \bibinfo{note}{{Paper 03-4266}}.

\bibitem[{\citenamefont{Williams}(1997)}]{WILLIAMS:97}
\bibinfo{author}{\bibfnamefont{J.~C.} \bibnamefont{Williams}}, in
  \emph{\bibinfo{booktitle}{Traffic Flow Theory: A State-of-the-Art Report}},
  edited by \bibinfo{editor}{\bibfnamefont{N.}~\bibnamefont{Gartner}},
  \bibinfo{editor}{\bibfnamefont{H.}~\bibnamefont{Mahmassani}},
  \bibinfo{editor}{\bibfnamefont{C.~J.} \bibnamefont{Messer}},
  \bibinfo{editor}{\bibfnamefont{H.}~\bibnamefont{Lieu}},
  \bibinfo{editor}{\bibfnamefont{R.}~\bibnamefont{Cunard}}, \bibnamefont{and}
  \bibinfo{editor}{\bibfnamefont{A.~K.} \bibnamefont{Rathi}},
  \bibinfo{organization}{Federal Highway Administration}
  (\bibinfo{publisher}{Transportation Reseach Board}, \bibinfo{year}{1997}),
  pp. \bibinfo{pages}{6.17--6.29}.

\bibitem[{\citenamefont{Lighthill and Whitham}(1955)}]{LIGHTHILL:55}
\bibinfo{author}{\bibfnamefont{M.}~\bibnamefont{Lighthill}} \bibnamefont{and}
  \bibinfo{author}{\bibfnamefont{G.}~\bibnamefont{Whitham}}, in
  \emph{\bibinfo{booktitle}{Proceedings of the Royal Society}}
  (\bibinfo{year}{1955}), vol. \bibinfo{volume}{A229}, pp.
  \bibinfo{pages}{317--345}.

\end{thebibliography}

\end{document}